\documentclass[11pt]{article}
\usepackage[utf8]{inputenc}
\usepackage[english]{babel}
\usepackage{amsmath}
\usepackage{amsthm}
\usepackage{graphicx}
\usepackage[margin = 1in]{geometry}
\usepackage[colorinlistoftodos]{todonotes}
\usepackage{algorithm}
\usepackage{algpseudocode}
\usepackage{amssymb}
\usepackage{subcaption}

\usepackage{natbib}
\usepackage{setspace}

\usepackage{amsfonts}

\usepackage{bm}

\newcommand{\cm}[1]{\ignorespaces}

\usepackage{booktabs,multirow,tabularx}
\usepackage{graphics}
\usepackage{graphicx}

\usepackage{xcolor}
\definecolor{mypink}{RGB}{219, 48, 122}

\definecolor{mypurple}{RGB}{75,0,130}

\usepackage{hyperref} 

\usepackage{amsmath}
\usepackage{amsthm}
	\newtheorem{theorem}{Theorem}
	 
	\newtheorem{proposition}{Proposition}

	\newtheorem{assumption}{Assumption}
	
\newcommand{\cS}{\mathcal{S}}
\newcommand{\cE}{\mathcal{E}}
\newcommand{\bE}{\mathbb{E}}
\def\bftheta{\boldsymbol \theta}
\def\bfeta{\boldsymbol \eta}
\def\bfnu{\boldsymbol \nu}
\def\bfmu{\boldsymbol \mu}
\def\bfzero{\boldsymbol 0}
\newcommand{\br}[1]{\left\{ #1 \right\} }
\newcommand{\mbr}[1]{\left[ #1 \right] }
\newcommand{\sbr}[1]{\left( #1 \right) }
\def\R{\mathbb R}
\def\bfC{\mathbf C}
\def\bfM{\mathbf M}
\def\bfm{\mathbf m}
\def\bfZ{\mathbf Z}
\def\bfX{\mathbf X}
\def\bfY{\mathbf Y}
\def\bfW{\mathbf W}
\def\bfN{\mathbf N}
\def\bfI{\mathbf I}
\def\bfb{\mathbf b}
\def\bfU{\mathbf U}
\def\cGP{\mathcal{GP}}
\def\rT{\mathrm T}
\def\dd{\text{d}}
\def\Var{\text{Var}}
\def\tM{\widetilde{M}} 
\def\SP{\text{SP}} 
\def\leb{\lambda}
\def\iid{\scriptsize \mbox{iid}}
\def\ind{\scriptsize \mbox{ind}}
\newcommand{\bias}{\text{bias}}
\newcommand{\pscale}{p}
\newcommand\numberthis{\addtocounter{equation}{1}\tag{\theequation}}

\usepackage{setspace}
\begin{document}

\title{Bayesian Structured Mediation Analysis With Unobserved Confounders}

\author{Yuliang Xu$^{1}$, 
Shu Yang$^{2}$ 
Jian Kang$^{3}$ \\
$^{1}$Department of Statistics, University of Chicago, Chicago, IL 60637, U.S.A.\\
$^{2}$Department of Statistics, North Carolina State University, Raleigh, NC 27695, U.S.A.\\
$^{3}$Department of Biostatistics, University of Michigan, Ann Arbor, MI 48109, U.S.A.}

\date{}
\maketitle


\begin{abstract}
We explore methods to reduce the impact of unobserved confounders on the causal mediation analysis of high-dimensional mediators with spatially smooth structures, such as brain imaging data. The key approach is to incorporate the spatial latent subject-specific spatial confounding effects, which influence the structured mediators, as unobserved confounders in the outcome model, thereby potentially debiasing the mediation effects. We develop BAyesian Structured Mediation analysis with Unobserved confounders (BASMU) framework, and establish its model identifiability conditions. Theoretical analysis is conducted on the asymptotic bias of the Natural Indirect Effect (NIE) and the Natural Direct Effect (NDE) when the unobserved confounders are omitted in mediation analysis. For BASMU, we propose a two-stage estimation algorithm to mitigate the impact of these unobserved confounders on estimating the mediation effect. Extensive simulations demonstrate that BASMU substantially reduces the bias in various scenarios. We apply BASMU to the analysis of fMRI data in the Adolescent Brain Cognitive Development (ABCD) study, focusing on four brain regions previously reported to exhibit meaningful mediation effects. Compared with the existing image mediation analysis method, BASMU identifies two to four times more voxels that have significant mediation effects, with the NIE increased by 41\%, and the NDE decreased by 26\%.
\end{abstract}

\noindent%
\vfill

\newpage

\section{Introduction}
\label{sec:intro}

High-dimensional mediation analysis is increasingly important in causal inference with complex data, especially as brain imaging and connectome datasets grow \citep{Lindquist2012-oj, chen2022imaging}. We propose a causal mediation framework that accounts for unobserved confounding in high-dimensional structured mediators, motivated by brain imaging data. Using fMRI data from the Adolescent Brain Cognitive Development (ABCD) study, we investigate how parental education affects children’s cognitive ability by identifying the neural pathways mediating this causal link.

Despite recent advances in imaging mediation analysis, the influence of unobserved confounders has largely been ignored. The no-unobserved-confounding assumption means that there exists no unobserved confounding variable that can impact both the outcome and the mediator. Existence of such unobserved confounders violates the sequential ignorability assumption proposed in \cite{Imai2010-ih}, where the predictor (treatment) is fully randomized but the mediator is not. Many recent advances in the high-dimensional mediation analyses rely on the no-unobserved-confounding assumption \citep{xu2023bayesian,Lindquist2012-oj, Nath2023-mv, song2020cors,jiang2023causal,wang2023high}.


The no-unmeasured-confounding assumption is ubiquitous in causal mediation analysis and general causal inference. Existing methods to address bias from unobserved confounders fall into two main categories: sensitivity analysis and external information (negative control or instrumental variable).
The sensitivity analysis approach bounds the Natural Indirect and Direct Effects (NIE, NDE) given assumptions about the scale of unobserved confounding \citep{Cinelli2020-ai,Zhang2022-yb,Imai2010-ih,Tchetgen2012-hf,ding2016sharp}. Classical approaches include parametric sensitivity analysis under linear structural equation models with scalar mediators \citep{Imai2010-ih} and nonparametric extensions \citep{Tchetgen2012-hf}; see \cite{Tchetgen2012-hf} for detailed comparisons. Others provide sharp bounds for binary treatment/outcome \citep{ding2016sharp} or use partial $R^2$ to quantify omitted-variable bias \citep{Cinelli2020-ai,Zhang2022-yb}.
However, sensitivity analysis cannot directly identify hidden confounders or produce debiased mediation effects. With high-dimensional mediators, it further requires specifying infinitely many possible combinations of confounder effects across spatial locations, making this approach especially restrictive. Approaches using instrumental variables (IVs) \citep{angrist1996identification,rudolph2024using} or negative controls \citep{lipsitch2010negative,miao2024confounding} rely on stringent validity conditions: IVs must affect the exposure (or mediator) without directly influencing the outcome and be independent of all hidden confounders; negative controls must share the same confounding structure while having no causal effect. Such variables are rarely available, particularly with complex neuroimaging mediators.
Other directions, such as using shared structure between confounders and exposure \citep{guan2023spectral}, lie outside the mediation framework.

Beyond bias from unobserved confounders, brain imaging mediation faces unique challenges: fMRI data are high-dimensional and spatially structured, with active mediation effects confined to small regions. The goal is to identify active mediation areas in the brain after accounting for unobserved confounding. Existing high-dimensional methods assume no unobserved confounding, and standard remedies (sensitivity analysis, instrumental variables, negative controls) do not adapt, leaving no principled way to adjust for spatially varying confounding.

To close this gap, we introduce BASMU (Bayesian Structured Mediation analysis with Unobserved confounders), a Bayesian framework that leverages spatial information to partially recover unidentifiable confounding effects. BASMU targets structured unobserved confounders with spatially smooth effects on the mediator, and directly estimates the subject-specific spatial confounding effect. Building on the classical linear structural equation model (LSEM) of \cite{Baron1986-yh}, we extend it to include unobserved confounders in both mediator and outcome models. Under mild conditions, these confounders can be estimated from the mediator model and used to debias the outcome model. We provide theoretical results on the asymptotic bias when such confounders are omitted and propose several algorithms to estimate subject-specific spatial confounding effects. BASMU allows flexible priors for the confounder coefficients and consistently outperforms competing methods in simulations, while increasing detection of significant mediation effects in ABCD brain imaging data.

Section 2 introduces the structured mediation framework with unobserved confounders and its identifiability and bias analysis. Section 3 outlines the two-stage estimator, Section 4 compares BIMA and BASMU in simulations, Section 5 applies the method to ABCD data, and Section 6 concludes.

\section{Bayesian Structured Mediation with Unobserved Confounders Framework}\label{sec:method}

Structured mediators are multivariate mediators with latent correlation, such as 2D/3D fMRI images, time series, or network data. For subject $i$, denote the functional mediator as $M_i(s),s\in\cS\subseteq \R^d$, where $\cS$ may be a spatial grid, time points, or a compact 3D brain domain. The mean function $\mathbb{E}\br{M_i(s)}$ is assumed to vary smoothly, for example through low-rank spatial bases or autoregressive time dependence. A scalar mediator with no variation in $\cS$ cannot separate unobserved confounding from spatial-independent noise. In structured mediation analysis, ignoring unobserved confounders yields biased estimates. Unlike approaches relying on sensitivity analysis or external instruments, we estimate unobserved confounders directly by exploiting spatial structure. Motivated by the ABCD study of children’s brain fMRI, we examine how parental education influences children’s general cognitive ability through the mediation of brain development at voxel locations $s_j\in\cS$.

For subject $i = 1,\dots,n$ and grid location $j=1,\dots,p$, let $M_i(s_j)$ be the image intensity, $X_i$ the scalar exposure, $\bfC_i \in \R^q$ the observed confounders, and $Y_i$ the scalar-valued outcome. The domain $\cS$ is partitioned into disjoint pixels $\br{\Delta s_j}_{j=1}^p$ of of equal Lebesgue measure $p^{-1}$ with centers $s_j$. We abbreviate grid-based functional values as $\eta_i = \br{\eta_i(s_j)}_{j=1}^p$, and similarly for $\alpha, \beta, \xi_k$. Standard notation includes $\cGP(0,\kappa)$ for a Gaussian Process with mean  0 and covariance $\kappa(\cdot,\cdot)$, and $|A|$ be the cardinality of a set $A$ (see Table \ref{tb:notation}).

\begin{figure}[htbp]
    \centering
    \includegraphics[width = 0.5\textwidth]{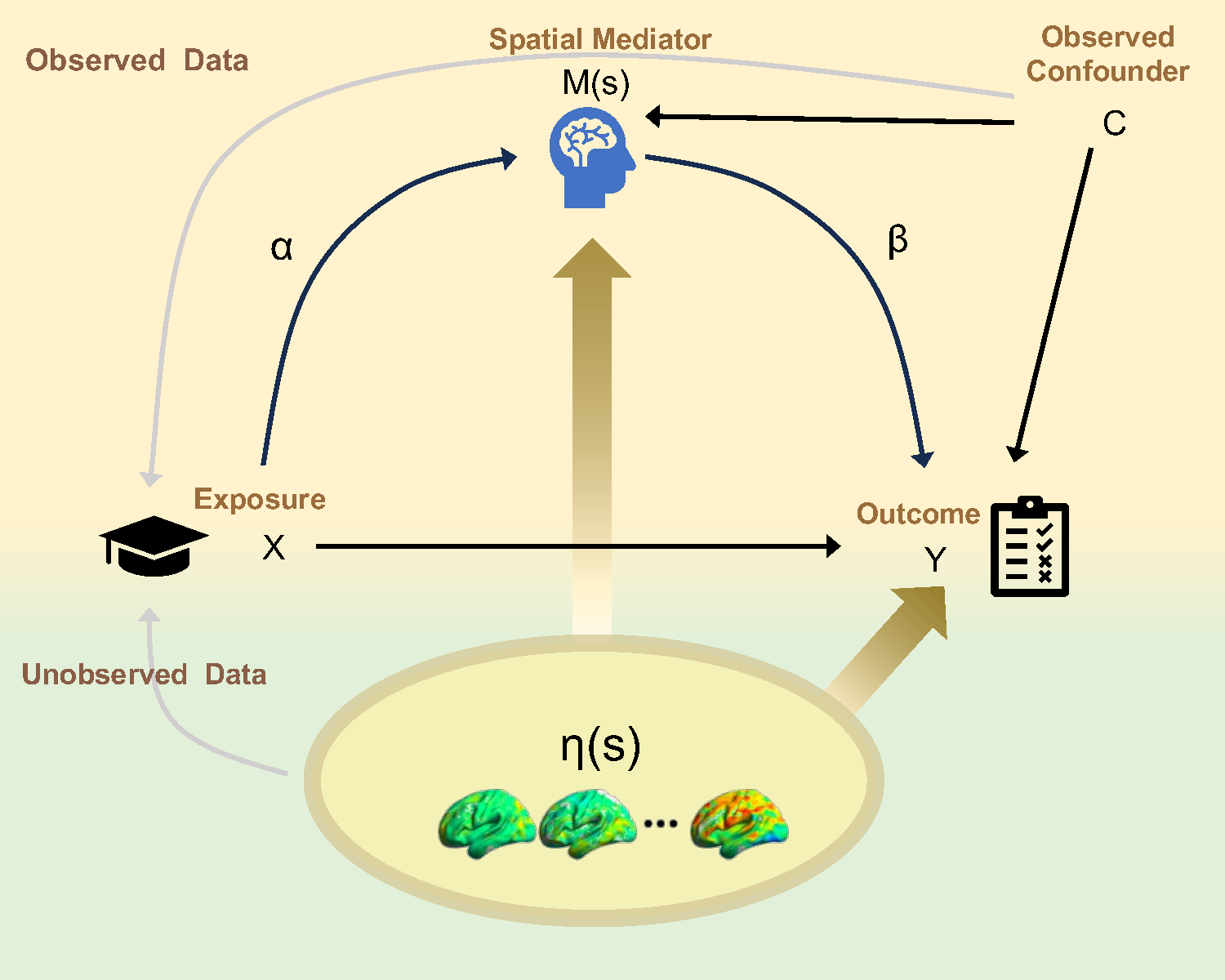}
    \caption{\textbf{Model Overview. } Causal graph representation of BASMU. $X\rightarrow Y$ is the Natural Direct Effect (NDE), and the mediation pathway $X\rightarrow M \rightarrow Y$ is the Natural Indirect Effect (NIE). The unobserved confounders may take on many forms. BASMU aims to account for the spatially varying unobserved confounders $\eta(s)$. The brown arrows represent the unobserved confounding effect ignored by many high-dimensional mediation methods. The grey arrows indicate we allow the confounders $\bfC_i,\eta_i$ to impact $X_i$, although this impact is not explicitly modeled (see details in the sensitivity analysis in Section~\ref{supp_sec:SA_causal}).}
    \label{fig:overview}
\end{figure}

We present the BASMU framework, based on the structural equation models. Figure \ref{fig:overview} provides the causal illustration.

\begin{align}
&M_i(s_j) = \alpha(s_j) X_i + \sum_{k=1}^q \xi_k(s_j) C_{i,k} + \eta_i(s_j)+ \epsilon_{M,i}(s_j), \label{eq:mediator}\\
&Y_i  =\sum_{j=1}^p \beta(s_j)M_i(\Delta s_j)  + \gamma X_i + \sum_{k=1}^q\zeta_k C_{i,k} + \sum_{j=1}^p \nu(s_j)\eta_i(s_j)\leb(\Delta s_j) + \epsilon_{Y,i},  \label{eq:full}
\end{align}
where $\epsilon_{Y,i}\overset{\iid}{\sim} N(0,\sigma^2_Y)$, $\epsilon_{M,i}(s_j)\overset{\iid}{\sim} N(0,\sigma^2_M)$. In model \eqref{eq:full}, $M_i(\Delta s_j) = \bE\br{M_i(s_j)}\leb(\Delta s_j) + \epsilon_{M,i}(\Delta s_j)$ with $\epsilon_{M,i}(\Delta s_j)\overset{\ind}{\sim} N(0,\sigma_M^2 \leb(\Delta s_j))$.

In the mediator model \eqref{eq:mediator},  we use $\alpha$ to denote the impact of the treatment $X_i$ on the image mediator $M_i$, $\xi_k$ the functional-coefficient for the vector-valued confounders $C_i\in\R^q$, $\eta_i$ the subject-specific spatial confounding effect, and $\epsilon_{M,i}$ the spatially-independent noise term. Model \eqref{eq:mediator} is the structured mediator where the mean part of \eqref{eq:mediator} follows a certain spatial pattern induced by Gaussian Process Kernels (see details for the GP priors in Section $\mathsection$\ref{sec:two_stage} and equation \eqref{eq:M_priors}).

In the outcome model \eqref{eq:full}, we use $\beta(s)$ to denote the mediator $M_i(\Delta s_j)$'s effect on the outcome $Y_i$. Here, $M_i(\Delta s)$ denotes the integrated effect of the random function $M_i(s)$ over a small voxel $\Delta s$. We use $\gamma,\zeta_k$ to denote the effect of the exposure $X_i$ and the observed confounder $C_{i,k}$, respectively. BASMU explicitly models unobserved confounders by exploiting spatial structure via the subject-specific spatial confounding effect $\eta_i(s)$, and uses the effect of $\eta_i(s)$ on $Y_i$ to represent the impact of unobserved confounders, denoted as $\nu(s)$.

The key intuition behind BASMU is that when an unobserved confounder influences both the mediator and the outcome, its effect on the mediator is unlikely to be homogeneous. Because the mediator itself exhibits spatial variation, the unobserved confounder must also leave a spatially varying footprint on the mediator. This latent structural effect can be captured by the subject-specific spatial confounding effect $\eta_i(s)$. By incorporating $\eta_i(s)$ into the outcome model, BASMU captures the potential influence of unobserved confounders on the outcome, thereby reducing bias that arises when such confounding is ignored.

Consider the following example: if there exists unobserved confounder $\bfZ_i$ of unknown dimension $m$, in the mediator model \eqref{eq:mediator}, there would be a term $\sum_{r=1}^m\xi_{r,z}(s_j)Z_{r,i}$ which plays the same role as $\sum_{k=1}^q \xi_k(s_j) C_{i,k}$. Since $\bfZ_i$ is unobservable, if $\xi_{m,z}(s_j)$ is a spatially-varying effect, we can replace the term $\sum_{r=1}^m\xi_{m,z}(s_j)Z_{r,i}$ by the subject-specific spatial confounding effect $\eta_i(s_j)$, as long as they have the same spatially-varying structure so that  $\eta_i(s_j)$ is separable from the independent noise $\epsilon_{M,i}(s_j)$.

With the presence of unobserved confounders $\eta_i$ as shown in the BASMU model  \eqref{eq:mediator} and \eqref{eq:full}, we follow the mediation assumption proposed in \cite{VanderWeele2014-hj} and make the following causal identification assumptions: for any $i$, endogenous $x$ and $\bfm$,  
\begin{align*}
    &\text{(i) } Y_{i,(x,\bfm)}\perp X_i\mid \br{\bfC_i,\eta_i},\ \text{(ii) }Y_{i,(x,\bfm)}\perp \bfM_i\mid \{\bfC_i,X_i, \eta_i\}, \\
    &\text{(iii) }\bfM_{i,(x)}\perp X_i\mid \br{\bfC_i,\eta_i}, \ \text{(iv) }Y_{i,(x,\bfm)} \perp \bfM_{i,(x')}\mid \br{\bfC_i, \eta_i }.\numberthis \label{eq:unobserved_assumption}
\end{align*}

The first three assumptions ensure that the confounders $C_i,\eta_i$ controls the outcome-treatment confounding, the outcome-mediator confounding, and the mediator-treatment confounding respectively. The fourth assumption ensures the outcome-mediator confounders are not affected by the underlying endogenous treatment $x$. For Assumption (iv) to hold, in the true data-generating mechanism, $\eta_i$ is allowed to be a baseline variable, but not allowed to be a post-exposure variable affected by $X_i$, that also affect both the mediator $M_i(s)$ and outcome $Y_i$,  while not being part of the mediator of interest. We include a detailed discussion on the implications of Assumption (iv) in Supplementary Section~\ref{supp_sec:additional_limitation}.\label{sent:main_causal}

The set of assumptions in \eqref{eq:unobserved_assumption} is the same as in \cite{ding2016sharp}, where a sensitivity analysis strategy is proposed for binary outcomes with binary exposures and scalar mediators. Here, we take advantage of the subject-specific spatial confounding effect $\eta_i$ as the unobserved confounders.
Under SUTVA (Stable Unit Treatment Value Assumption) and causal identification assumptions in \eqref{eq:unobserved_assumption}, following a similar argument in \cite{Lindquist2012-oj}, we can derive the NIE and NDE in the form of estimable parameters using the proposed structural equation models~\eqref{eq:mediator} and \eqref{eq:full}.

Conditional on $\br{\bfC_i,\eta_i}$, the average treatment effect between $x$ and $x'$ is $$\bE\left[ Y_{i,\{x, \bfM_{i,(x)}\}} - Y_{i,\{x', \bfM_{i,(x')}\}} \right],$$ which can be decomposed into the NIE, $$\bE\left[ Y_{i,\{x, \bfM_{i,(x)}\}} - Y_{i,\{x, \bfM_{i,(x')}\}} \right] = \sum_{j=1}^{\pscale}  \beta(s_j)\alpha(s_j)\leb(\Delta s_j)(x-x'),$$ and the NDE, $$\bE\left[  Y_{i,\{x, \bfM_{i,(x')}\}} - Y_{i,\{x', \bfM_{i,(x')}\}}\right] = \gamma(x-x').$$
 Our primary interest is in estimating the spatially-varying NIE, defined as $\mathcal{E}(s) = \alpha(s)\beta(s)$, and the scalar-valued NIE is $\mathcal{E} = \sum_{j=1}^p \alpha(s_j)\beta(s_j)\leb(\Delta s_j)$. 
 For spatial mediation analysis, we are not only interested in the scalar-valued NIE $\cE$, but also the spatial decomposition of NIE in $\cS$ where $\cE(s)$ is nonzero.

The joint identifiability of models \eqref{eq:mediator} and \eqref{eq:full} is non-trivial, especially with the introduction of $\eta_i$ and $\nu$. We impose a set of model identifiability assumptions.

Define column vectors $\bfX = \sbr{X_1,\dots, X_n}^\rT\in \R^{n}$, $\bfC = \sbr{\bfC_1,\dots, \bfC_n}^\rT\in\R^{n\times q}$. Let $\br{\psi_l(s)}_{l=1}^\infty$ be a set of basis of $L^2(\cS)$. Denote the basis coefficients $\theta_{\eta_i,l} = \int_{\cS} \eta_i(s)\psi_l(s)\leb(\dd s)$, and $ \bftheta_{\eta,l} = \sbr{\theta_{\eta_i,l},\dots, \theta_{\eta_n,l} }$. 

\begin{assumption}\label{asm:iden}
(i) Conditional on $(\bfX,\bfC)$, there exists a constant matrix $\bfW = (W_{i,k})\in \R^{n\times (q+1)}$ such that $\det\{\bfW^\top (\bfX,\bfC)\}\neq 0$;  (ii) There exists a constant vector $\bfb=(b_1,\ldots, b_q)^\top $ such that for any $s\in\cS$ and $k=1,\ldots,q+1$,  $\sum_{i=1}^n W_{i,k} \eta_i(s) = b_k$; (iii) The design matrix after basis decomposition $\br{\bfX,\bfC, \bftheta_{\eta,1},\dots ,\bftheta_{\eta,L}}\in \R^{n\times (L+1+q)}$ is assumed to be full rank, and for any subset $\cS_m\subset \cS$ where $|\cS_m|=m$, the design matrix before basis decomposition $\br{\bfX,\bfC, \br{\bfeta(s_k)}_{s_k\in\cS_m}}\in \R^{n\times (m+1+q)}$ is also assumed to be full rank. 
(iv) The unobserved confounding effect $\nu$ is either low-rank (i.e. $\nu(s) = \sum_{l=1}^L\theta_{\nu,l}\psi_l(s)$, $L=o(n)$), or is sparse (i.e. $\bfnu=(\nu(s_1),\dots,\nu(s_p))\in \br{v\in \R^p: \|v\|_0=m}$, $m=o(n)$).
\end{assumption}

\textit{Remark.}
Assumption \ref{asm:iden} is to guarantee the identifiability of the BASMU model. Assumption \ref{asm:iden} (i) and (ii) are used in the proof of identifiability of $\br{\eta_i}_{i=1}^n$ in the mediator \eqref{eq:mediator}, which guarantees the identifiability of all parameters in the mediator model \eqref{eq:mediator}. Assumption \ref{asm:iden} (iii) and (iv) are to ensure that the design matrix in the outcome model \eqref{eq:full} is full-rank and $\nu$ is either sparse or low-rank, which ensures the identifiability of $\beta, \gamma, \br{\zeta_k}_{k=1}^q,\nu,\sigma_Y$ in the outcome model \eqref{eq:full} given $\br{\eta_i}_{i=1}^n$. One example of Assumption \ref{asm:iden} is  $\bfb=0$ and $\bfW = (\bfX,\bfC)$, and $\bftheta_{\eta,l}\in \R^n$ are sampled from a subspace of $\R^n$ orthogonal to $\text{span}\br{\bfX, \bfC}$. Note that this orthogonality is in the sense of linear algebra, meaning that there are $n-1-q$ free random variables in $\bftheta_{\eta,l}$. This does not mean $\bftheta_{\eta,l}$ is independent from $\br{\bfX, \bfC}$ as random variables; on the contrary, this assumption restricts the sample space of $\bftheta_{\eta,l}$ to be constrained by $\br{\bfX, \bfC}$.

Let $\bftheta_{\text{all}} = \{\alpha,\br{\xi_k}_{k=1}^q, \br{\eta_i}_{i=1}^n, \sigma_M, \beta, \gamma, \br{\zeta_k}_{k=1}^q,\nu,\sigma_Y\}$ be the collection of all parameters in model \eqref{eq:mediator} and \eqref{eq:full}.
\begin{proposition}\label{prop:identifiable}
Under Assumption \ref{asm:iden}, the BASMU model in \eqref{eq:mediator} and \eqref{eq:full} is jointly identifiable, i.e. given density function $\prod_i f(Y_i, \bfM_i;\bftheta| X_i, \bfC_i)$, 
$$\prod_i f(Y_i, \bfM_i;\bftheta_{\text{all}}| X_i, \bfC_i) = \prod_i f(Y_i, \bfM_i;\bftheta^*_{\text{all}}| X_i, \bfC_i)$$ implies $\bftheta_{\text{all}} = \bftheta^*_{\text{all}}$. 
\end{proposition}

Proposition \ref{prop:identifiable} shows that as long as $\nu$ has a latent low-dimensional representation, or $\nu$ is sparse, the proposed BASMU model \eqref{eq:mediator} and \eqref{eq:full} are jointly identifiable. In the next section, we analyze the bias induced by ignoring the unobserved confounder in \eqref{eq:full}.


\subsection{Structured Mediation When Omitting Unobserved Confounders}\label{subsec:omit_uc}

If we ignore the impact of unobserved spatial confounder on the outcome, the potentially biased outcome model can be defined as
\begin{align*}
     &Y_i  =\sum_{j=1}^p \beta(s_j)M_i(\Delta s_j)  + \gamma X_i + \sum_{k=1}^q\zeta_k C_{i,k} + \epsilon_{Y,i}, \quad\epsilon_{Y,i}\overset{i.i.d}{\sim} N(0,\sigma^2_Y).  \numberthis \label{eq:reduced}
\end{align*}

This outcome model \eqref{eq:reduced}, a simplified special case of our proposed BASMU model when $\nu(s)\equiv 0$ for all $s\in \cS$, is widely applied in many high-dimensional mediation analysis methods \citep{Lindquist2012-oj,song2020cors,song2021ptg,jiang2023causal,wang2023high,xu2023bayesian}. These methods often assume the high dimensional mediation effects have some low-dimensional latent structures; the difference is in the various low-dimensional space representations, including but not limited to: mapping through a known set of bases, such as the piecewise polynomials B-spline basis \citep{Lindquist2012-oj}, or the global Bernstein polynomial basis \citep{jiang2023causal}, in the Bayesian alternative, through Gaussian Process (GP) priors \citep{xu2023bayesian} or vector-valued priors with a dense covariance matrix \citep{song2020cors}, or imposing a conditional autoregressive correlation structure \citep{wang2023high}. Using low-dimensional representations to approximate the high-dimensional parameter is an effective and ubiquitous approach in many high-dimensional data analysis problems. This makes our proposed BASMU framework even more valuable, because this idea can essentially be extended to any type of low-dimensional representations simply by changing the structure of $\eta_i$ to the corresponding low-dimensional manifold. 
    As one example, Bayesian Image Mediation Analysis (BIMA) \citep{xu2023bayesian} has demonstrated the flexibility of using GP priors since the low-dimensional representation can be adjusted using different covariance kernels. We refer to model \eqref{eq:mediator} and \eqref{eq:reduced}  as the Bayesian Image Mediation Analysis (BIMA) framework \citep{xu2023bayesian}, where the unobserved confounders are omitted.

\subsection{Bayesian Bias Analysis With Omitted Unobserved Confounders}

The bias analysis highlights the need for BASMU: spatial mediation models like BIMA are unbiased only without unobserved confounders, but in practice, bias is inevitable and depends on both confounder scale and spatial mediator distribution.
The posterior mean of $\alpha$ in the mediator model \eqref{eq:mediator} is a consistent estimator \citep{xu2023bayesian}. Hence the main focus is on the asymptotic bias of the posterior mean of $\beta$ under \eqref{eq:reduced} as a point estimator. We assign a GP prior on $\beta\sim \cGP(0,\sigma_\beta^2\kappa)$. By Mercer's theorem, for a given kernel $\kappa(s,s') = \sum_{l=1}^\infty\lambda_l \psi_l(s)\psi_{l}(s')$, we can represent $\beta(s) =  \sum_{l=1}^L\theta_{\theta_\beta,l} \psi_l(s)$, $L=o(n)$ where $\theta_{\beta,l} \overset{\ind}{\sim} N(0,\lambda_l\sigma_\beta^2)$.

We use a superscript `0' to denote a true parameter value, e.g., $\theta^0$. Denote the true unobserved confounder term $U^0 := \sbr{\bfeta^0}^\rT\nu^0 \in \R^{n}$, and given estimators $\widehat\bfeta$ and $\widehat\nu$, denote $\widehat U := \sbr{\widehat\bfeta}^\rT\widehat\nu \in \R^{n}$. The result below discusses the asymptotic bias of $\beta$ under model \eqref{eq:reduced} where the unobserved confounder is omitted,  and model \eqref{eq:full} where the unobserved confounder is considered.

For the structured mediator $M_i(\Delta s_j) = \bE\br{M_i(s_j)}\leb(\Delta s_j) + \epsilon_{M,i}(\Delta s_j)$, $\epsilon_{M,i}(\Delta s_j)\overset{\ind}{\sim} N(0,\sigma_M^2 \leb(\Delta s_j))$, denote $E_i(s) := \bE\br{M_i(s)}$. In addition, denote the basis coefficients under GP basis decomposition as $\theta_{E,i,l} = \int_{\cS} E_i(s)\psi_l(s)\leb(\dd s)$, where $\psi_l$ is the same basis function in the GP prior of $\beta$.

\begin{assumption}\label{asm:M_mean}
     (i) For any $l,l'$, $\lim_{n\to\infty}n^{-1}\sum_{i=1}^n \theta_{E,i,l} \theta_{E,i,l'} = H_{l,l'}$, where $H_{l,l'}$ is some finite constant; 
     (ii) For any $l$, $\lim_{n\to\infty} n^{-1}\sum_{i=1}^n \theta_{E,i,l} U_i^0 = h^0_l$, where $h^0_l$ is a finite constant, $U_i^0$ is the $i$-th element in $U^0$.
\end{assumption}

\begin{assumption}\label{asm:h_hat}
Conditional on $\widehat\bfeta$ and $\widehat \nu$, $\lim_{n\to\infty} n^{-1}\sum_{i=1}^n \theta_{E,i,l} \widehat U_i \overset{p}{\to} \widehat h_l$ where $\widehat h_l$ is a random variable that only depends on $\widehat\bfeta$ and $\widehat \nu$.
\end{assumption}

One example of Assumption \ref{asm:M_mean} is to view $E_i(s)$ as i.i.d samples from some unknown process $E(s)$ (for example, Gaussian Process) with finite first and second moments, and $H_{l,l'}$ is the finite covariance for the basis coefficients at $l$ and $l'$. If we view elements in $U^0$ as n i.i.d samples drawn from a distribution $U$ with finite second moments, $h_l^0$ is the covariance between the $l$-th basis coefficient of $E(s)$ and $U$, and is also finite (Holder's inequality). The same example applies to Assumption \ref{asm:h_hat}.

In the Theorem \ref{prop:bias} below, we use $\widehat\theta_\beta^B$ to denote the point estimator under the BIMA outcome model and use $\widehat\theta_\beta^F$ to denote the point estimator under the full BASMU outcome model that takes account into the unobserved confounders. Denote $\widetilde M_{i,l} = \int_{S}M_i(s)\psi_l(s)\leb(\dd s)$ and $\tM\in \R^{n\times L}$, $(\tM)_{i,l} = \widetilde M_{i,l}$, and denote $A := \tM^\rT\tM\in\R^{L\times L}$ where $L=o(n)$. Denote $\sigma_{\min}(A)$ and $\sigma_{\max}(A)$ as the smallest and largest singular values of $A$ respectively.

\begin{theorem}\label{prop:bias}
     Assume that $A$ satisfy $$0<c_{\min}<\lim\inf_{n\to\infty}\sigma_{\min}(A)/n \leq \lim\sup_{n\to\infty}\sigma_{\max}(A)/n\leq c_{\max}<\infty$$ with probability $1-\exp\br{-c_0n}$ for some positive constant $c_0, c_{\min}, c_{\max}$. In addition, in the GP prior basis coefficients $\theta_{\beta,l}\overset{\ind}{\sim}N(0,\sigma_\beta^2\lambda_l)$ , assume that $\lambda_L > c_{\lambda} n^{-1+a_\lambda}$  for some positive constant $c_\lambda,a_\lambda$. We can draw the following conclusions given Assumptions \ref{asm:M_mean} and \ref{asm:h_hat}:
    
    (i) When $U^0=\bf0$, i.e. no unobserved confounder, the asymptotic bias of $\widehat\theta_\beta^B$ is 0.

    (ii) Given Assumptions \ref{asm:M_mean} and \ref{asm:h_hat}, and assume that the true unobserved confounder term $\sum_{i=1}^n\sbr{ U_i^0}^2/n$ is finite,  then the bias of the posterior mean of $\theta_\beta$ under BIMA model \eqref{eq:reduced} $\bias(\widehat\theta_\beta^B ) \overset{p}{\to} (H+\sigma_M^2 I_L)^{-1}h^0$, and the bias under the full model \eqref{eq:full} $\bias(\widehat\theta_\beta^F ) \overset{p}{\to} (H+\sigma_M^2 I_L)^{-1}\sbr{h^0-\widehat h}$.
\end{theorem}

\textit{Remark.} The result of Theorem \ref{prop:bias} is conditional on the true values of $\gamma$ and $\zeta_k$ in \eqref{eq:full} for simplicity of the analysis. A similar bias analysis result can be drawn on the NDE $\gamma$ if we treat $\bfX$ as one additional column in $\tM$, and the bias of $\gamma$ is the corresponding element in $\bias(\widehat\theta_\beta^B )$ and $\bias(\widehat\theta_\beta^F )$.

\textit{Implications.} Theorem \ref{prop:bias} (i) can be seen as a corollary of (ii). Based on Theorem \ref{prop:bias} (ii), we can expect that: (a) the bias of BIMA depends on the scale of the unobserved term $h^0$, and is nonzero unless $h^0=0$; (b) the bias of BASMU depends on the estimation of $\widehat\eta$ and $\widehat\nu$; (c) larger random noise $\sigma_M$ in the mediator model may reduce the bias, because larger random noise makes the observed mediator $M_i(s)$ less correlated with the subject-specific spatial confounding effect $\eta_i(s)$.

\section{Two-stage Estimation}\label{sec:two_stage}

The most natural way to estimate models \eqref{eq:mediator} and \eqref{eq:full} is to use a fully Bayesian approach to update all parameters iteratively. This involves updating the subject-specific spatial confounding effects $\eta_i$ jointly from both models \eqref{eq:mediator} and \eqref{eq:full} in every iteration. Due to the large parameter space of $\br{\eta_i}_{i=1}^n$ to search from, the joint estimation approach usually takes very long to converge. For every iteration when $\br{\eta_i}_{i=1}^n$ is updated, the new $\br{\eta_i}_{i=1}^n$ can have a huge impact on the likelihood of \eqref{eq:full}, and all other parameters in \eqref{eq:full} need longer iterations to converge to stable values, hence the joint estimation can make estimating the outcome model \eqref{eq:full} very unstable. Because the posterior of $\br{\eta_i}_{i=1}^n$ is mainly dominated by the mediator model \eqref{eq:mediator}, sampling $\br{\eta_i}_{i=1}^n$  based solely on \eqref{eq:mediator} can already give a consistent estimation (see remark of Assumption \ref{asm:iden}), hence instead of the fully Bayesian joint estimation approach, we propose the two-stage estimation.

In the two-stage estimation, we compute the posterior of model \eqref{eq:mediator} and \eqref{eq:full} separately. First, we draw posterior samples based on model \eqref{eq:mediator} based on the priors 
\begin{equation}
    \alpha \sim \cGP(0,\sigma^2_\alpha\kappa), \xi_k \sim \cGP(0,\sigma^2_\xi\kappa), \eta_i \sim \cGP(0,\sigma^2_\eta\kappa), \label{eq:M_priors}
\end{equation}
and compute the posterior mean of $\eta_i$ conditional only on model \eqref{eq:mediator}, denoted as $\widehat\eta_i$. Using $\widehat\eta_i$ as part of the fixed design matrix in \eqref{eq:full}, draw MCMC samples from \eqref{eq:full} conditioning on $\eta_i=\widehat\eta_i$, based on the following prior for $\nu$,
\begin{align*}
    \nu(s) &= g(s)\delta(s), \quad g(s)\overset{\ind}{\sim} N(0,\sigma_\nu^2),
    \quad \delta(s) \overset{\ind}{\sim} \text{Ber}(1/2). \numberthis\label{eq:prior_nu}
\end{align*}

The two-stage estimator uses a flexible spatially independent prior on $\nu$ with a sparsity-inducing variable $\delta$. Simulations show $\widehat\eta_i$ estimates $\eta_i$ well when $p\gg L$ (i.e., the $\eta_i$ kernel is smooth). Estimating $\nu$ remains challenging if $\eta_i$ is misspecified, but the flexible prior still greatly improves $\beta$ estimation over BIMA. For faster computation, we apply SVD to $\br{\eta_i}_{i=1}^n$; the full algorithm is in Web Algorithm \ref{algo:two_stage}.

We compare BIMA, which ignores unobserved confounders, with BASMU’s two-stage estimator in simulated examples. Though not fully Bayesian, it yields accurate point estimates and debiases $\beta$ provided $\widehat\eta_i$ is reasonably close to the truth. Supplementary Section~\ref{sec_supp:two_stage} provides additional simulation comparisons among different implementations of BASMU, including a fully joint update algorithm.

\section{Simulation Study}\label{sec:simulation}

\subsection{Simulation I: high-dimensional comparison under varying cases}

We compare the performance of  BIMA and BASMU through extensive simulation studies. For $\alpha$, $\beta$, and $\nu$, we simulate 2D $p= 40\times 40$ images (true signals are shown as in Web Figure \ref{fig:true_signal} and Web Figure \ref{fig:nu_pattern}). For the GP priors in \eqref{eq:M_priors} and $\beta$, we use the Mat\'ern kernel with $\rho=2,\tau=0.2, d=2$, 
\begin{align}\label{eq:matern}
    \kappa(s',s;\tau,\rho) = C_\nu(\|s'-s\|_2^2/\rho), ~ C_\tau(d):= \frac{2^{1-\tau}}{\Gamma(\tau)}\left( \sqrt{2\tau} d \right)^\tau K_\tau(\sqrt{2\tau d}).
\end{align}
The GP prior parameters $\alpha,\beta,\eta_i,\xi_k$ use the same basis decomposition of the kernel function in \eqref{eq:matern}, denoted as $\kappa(s',s) = \sum_{l=1}^\infty \lambda_l\psi_l(s')\psi_l(s)$. For example, $\beta(s)$ is approximated by $\sum_{l=1}^L\theta_{\beta,l}\phi_l(s)$ with the prior $\theta_{\beta,l}\overset{\ind}{\sim}N(0,\lambda_l\sigma_{\beta}^2)$.
Set $L=120$ basis coefficients as the cutoff. We use the Metropolis-Adjusted Langevin Algorithm (MALA) for updating $\alpha$ and $\beta$, and the Gibbs sampler for the rest of the parameters. For the outcome models in both BIMA and BASMU, we use a total of $2\times 10^4$ iterations with the last 10\% used as MCMC samples. The mediator model \eqref{eq:mediator} uses $10^3$ iterations with the last 10\% used as MCMC samples.

Table \ref{tb:sim_result} outlines six cases varying $\sigma_\eta$, $\sigma_M$, and $n$ to illustrate Theorem \ref{prop:bias}, with three $\nu$ signal patterns (dense, sparse, zero; Web Figure \ref{fig:nu_pattern}), based on 100 replications. Dense $\nu$ signals use low-dimensional basis coefficients mapped to $\R^p$ via the Mat\'ern kernel \eqref{eq:matern}. We report the bias, variance, and MSE of the scalar NIE $\cE$  (Table \ref{tb:sim_result}), and visualize voxelwise MSE and bias of the posterior mean $\beta(s)$ over 100 replications (Web Figures \ref{fig:Beta_MSE}, \ref{fig:Beta_bias}). Each voxel $s_j$ shows the bias/MSE of its posterior mean.

\begin{table}[ht!]
\centering
\caption{Simulation result of the scalar NIE $\cE$ averaged over 100 replications. The smaller MSE of $\cE$ is bolded in each case.  The default generative parameter settings are $\sigma_\eta=0.5$, $n=300$, $\sigma_M=2$.} 
\label{tb:sim_result}

\begin{tabular}{lcclcclcc}
\toprule
       & \textbf{BIMA}            & \textbf{BASMU}          &        & \textbf{BIMA}                & \textbf{BASMU}               &        & \textbf{BIMA}                    & \textbf{BASMU}                  \\
Case 1 & \multicolumn{2}{c}{dense $\nu$}  & Case 3 & \multicolumn{2}{c}{all 0 $\nu$}           & Case 5 & \multicolumn{2}{c}{dense $\nu$, $\sigma_\eta=1$} \\
Bias   & -2.72           & 1.06           & Bias   & -0.5                & -0.17               & Bias   & 13.31                   & 2.21                   \\
Var    & 3.59            & 4.11           & Var    & 2.31                & 4.49                & Var    & 3.35                    & 3.2                    \\
MSE    & 10.97           & \textbf{5.20 }          & MSE    & \textbf{2.53}                & 4.48                & MSE    & 180.36                  & \textbf{8.06}                   \\
Case 2 & \multicolumn{2}{c}{sparse $\nu$} & Case 4 & \multicolumn{2}{c}{sparse $\nu$, $n=600$} & Case 6 & \multicolumn{2}{c}{dense $\nu$, $\sigma_M=4$}    \\
Bias   & 7.56            & 1.87           & Bias   & 10.77               & 2.29                & Bias   & -6.33                   & -0.57                  \\
Var    & 3.42            & 3.76           & Var    & 1.54                & 1.54                & Var    & 13.43                   & 12.63                  \\
MSE    & 60.51           & \textbf{7.22  }         & MSE    & 117.5               & \textbf{6.79}                & MSE    & 53.36                   & \textbf{12.82}        \\
\bottomrule
\end{tabular}
\end{table}

Results in Table \ref{tb:sim_result} show that BASMU generally achieves the lowest MSE for $\cE$, except for Case 3 where no unobserved confounding effect is present. To verify the theoretical implications by Theorem \ref{prop:bias}, comparing Case 2 and 4, as $n$ increases, the MSE of BASMU decreases, whereas the BIMA model has increased MSE and bias. This is because $\|h^0-\widehat h\|_2$ decreases as $\widehat U \to U^0$ when $n$ increases. Comparing Case 1 and 5, as $\sigma_\eta$ increases, $U^0$ increases, the MSE for BASMU has little changes compared to the huge increase in MSE and bias for BIMA due to the increased scale of $U^0$. Comparing Case 1 and 6, as $\sigma_M$ increases, the bias for BASMU decreases. In fact, from the spatial MSE and bias in  Web Figure \ref{fig:Beta_bias} Case 6 compared to Case 1, both BIMA and BASMU have an overall decreased MSE and bias in $\beta$, though the decreased bias area does not overlap with the true nonzero signal regions in $\alpha$ and $\beta$, hence not fully reflected on the result of scalar NIE.

Web Figures \ref{fig:Beta_MSE} and \ref{fig:Beta_bias} show straight-forward evidence that the two-stage estimation of BASMU can indeed reduce the bias of $\beta(s)$ and have a lower MSE over varying spatial locations in all scenarios. In Web  Table \ref{tb:sim_result_NDE} we also provide the NDE result with similar implications.

\subsection{Simulation II: low-dimensional comparison across different mediation frameworks}\label{subsec:sim_lowd}

We compare BASMU with PTG \citep{song2021ptg}, HDMA \citep{gao2019testing}, and BIMA in a low-dimensional simulation ($p=32\times32$). Existing high-dimensional mediation models are affected differently in the presence of unmeasured confounders. Supplementary Section~\ref{supp_sec:describe_competing_methods} provides a brief description of PTG and HDMA.
We chose these three methods because each represents a different high-dimensional mediation framework. HDMA represents the Frequentist screening followed by de-biased estimation approach \citep{zeng2021statistical}, and is recommended as one of the best-performing methods for identifying active mediators in \cite{hdmed}. PTG represents the Bayesian style composite modeling of  $\alpha(s_j),\beta(s_j)$ with shrinkage priors. BIMA represents the Bayesian approach that models the low-dimensional structure of functional parameters with spatially smooth GP priors.

We use a low-dimensional case because PTG and HDMA are computationally intensive in higher dimensions. Table \ref{tb:sim_lowd} reports NIE and NDE point estimates, with posterior means for PTG, BIMA, and BASMU. BASMU achieves the lowest MSE for both NIE and NDE. Among methods assuming no unmeasured confounding, BIMA and HDMA have similar MSEs, but HDMA shows higher variance while BIMA is more biased. BIMA consistently captures the same confounded low-dimensional structure, while HDMA’s spatial-independent screening varies across datasets as unmeasured confounding $\eta_i$ changes.

\begin{table}[ht!]
\centering
\caption{Simulation result of the scalar NIE $\cE$ and NDE $\gamma$ averaged over 100 replications. The smallest MSE of $\cE$ is bolded in each case.  The default generative parameter settings are $\sigma_\eta=0.5$, $n=300$, $\sigma_M=2$, $p=32\times 32$. $\nu$ is the sparse triangle.} 
\label{tb:sim_lowd}
\begin{tabular}{llllllll}
\toprule
      & \multicolumn{3}{c}{\textbf{NIE}} & &\multicolumn{3}{c}{\textbf{NDE}} \\
      & Bias   & Var   & MSE    && Bias   & Var   & MSE    \\
\textbf{PTG}   & -10.55 & 4.54  & 115.77 &\textbf{PTG }  & -18.07 & 12.54 & 338.89 \\
\textbf{HDMA}  & -4.2   & 30.25 & 47.55  &\textbf{HDMA}  & 4.21   & 28.4  & 45.86  \\
\textbf{BIMA}  & 7.5    & 1.78  & 58.02  &\textbf{BIMA}  &  -6.26  & 0.19  & 39.33  \\
\textbf{BASMU} & 2.04   & 1.37  & \textbf{5.52}   &\textbf{BASMU} & -2.34  & 0.06  & \textbf{5.52}  \\
\bottomrule
\end{tabular}
\end{table}

\section{ABCD Data Application}\label{sec:real_data}

For the real-data analysis, we use ABCD release 1 \citep{casey2018ABCD} to study how children’s task fMRI signals mediate the effect of parental education (bachelor’s degree or higher) on children’s general cognitive ability. The structured mediator is 3D task fMRI data; confounders include age, gender, race/ethnicity (Asian, Black, Hispanic, Other, White), and household income ($<50$k, 50–100k, $>100$k), with multi-level variables binary coded. For the potential unobserved confounders, there are many known and unknown factors that could influence both the mediator (fMRI signal) and the outcome (children's cognitive score). We could only give a few examples that may have such spatial effects, including cognitive growth heterogeneity due to nutritional intake, differential exposure to neurotoxic pollutants (e.g., lead, air pollution) that alter brain development, regional socioeconomic disparities affecting educational opportunities and chronic stress levels, genetic polymorphisms influencing cortical maturation, and differences in sleep patterns or physical activity that modulate neural activation patterns.

The task fMRI data is 2-back 3mm task contrast data, and the preprocessing method is described in \cite{sripada2020toward}. Each voxel in the 3D image mediator represents the brain signal intensity when the subject tries to remember the tasks they performed 2 rounds ago. 
Using the same dataset as \cite{xu2023bayesian}, which ignored unobserved confounders, we reanalyze the four regions with the strongest mediation effects (Table \ref{tb:RDA_result}) using BASMU.
After preprocessing, we have $n=1861$ subjects and $p=3539$ voxels. The four regions of interest ($p=3539$ voxels) define the support $\cS$.. We use the same Mat\'ern GP kernel as \cite{xu2023bayesian}. BIMA and BASMU share the mediator model results with $10^4$ iterations; BIMA’s outcome model uses $3\times10^4$ iterations. For BASMU, we first fix $\delta(s)=1$ and run $10^4$ iterations to obtain initial values, then run the two-stage algorithm for $2\times10^3$ iterations, keeping the last 20\% as the MCMC sample. Runtime: 54 minutes (mediator), 61 minutes (BIMA outcome), 4 hours (BASMU outcome) on an Apple M1 laptop with 8 GB memory.

\begin{table}[ht]
\centering
\caption{Comparison of ABCD data analysis under BIMA and BASMU. The top table reports the active voxel selection, from column 3 to 8: number of active voxels selected by BIMA/BASMU (brackets: percentage of selected voxels over the total number of voxels), number of commonly selected voxels, number of voxels only selected by BIMA/BASMU, and the total number of voxels in each region. The bottom table reports the numeric values of the NIE, from column 3 to 8: summation of NIE over the region under BIMA/BASMU, summation of NIE over voxels with positive effect under BIMA/BASMU, summation of NIE over voxels with negative effect under BIMA/BASMU.}
\label{tb:RDA_result}
\resizebox{\columnwidth}{!}{
\begin{tabular}{lllllll}
\toprule
\multicolumn{7}{c}{\textbf{Selection of active mediation voxels $\cE(s_j)$}}                                                                                       \\
Region code and name               & BIMA     & BASMU     & common       & BIMA\_only    & BASMU\_only  & size          \\
34  Cingulum\_Mid\_R & 80 (13\%)             & \textbf{342 (57\%)}            & 68             & 12              & 274            & 605             \\
57  Postcentral\_L   & 108 (9\%)           & \textbf{246  (21\%)}           & 92             & 16              & 154            & 1159            \\
61  Parietal\_Inf\_L & 138  (20\%)          & \textbf{387  (56\%)}           & 131            & 7               & 256            & 696             \\
67  Precuneus\_L     & 150  (14\%)          & \textbf{404   (37\%)}          & 137            & 13              & 267            & 1079            \\

\hline
\multicolumn{7}{c}{\textbf{Effect size of $\cE$}}                                                                                     \\
Region code and name              & BIMA NIE & BASMU NIE & BIMA NIE (+) & BASMU NIE (+) & BIMA NIE (-) & BASMU NIE (-) \\
34  Cingulum\_Mid\_R & 0.007          & 0.007           & 0.022          & 0.032           & -0.015         & -0.025          \\
57  Postcentral\_L   & 0.007          & \textbf{0.021}           & 0.068          & 0.087           & -0.062         & -0.065          \\
61  Parietal\_Inf\_L & 0.009          & 0.000           & 0.163          & 0.163           & -0.154         & -0.162          \\
67  Precuneus\_L     & 0.051          & 0.075           & 0.107          & 0.140           & -0.057         & -0.065         
     \\
\bottomrule
\end{tabular}
}
\end{table}

\begin{figure}[ht]
    \centering
    \includegraphics[width = 0.7\textwidth]{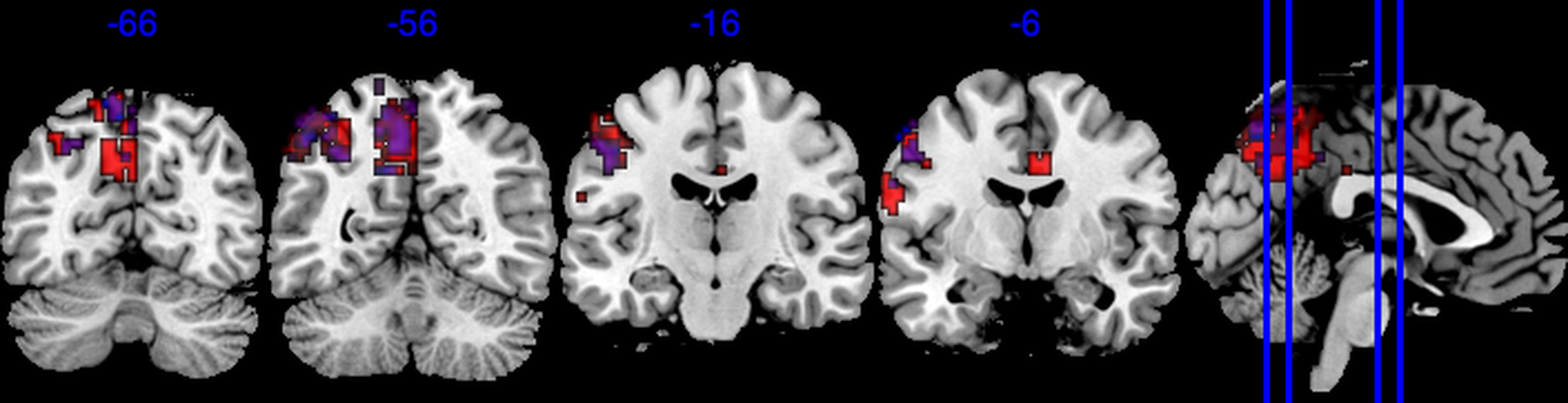}
    \caption{Coronal view of active NIE $\cE$ areas. The blue areas are active mediation voxels selected by BIMA, the red areas are selected by BASMU, and the overlaying purple areas are commonly selected by both methods.}
    \label{fig:RDA_TIE}
\end{figure}

Table \ref{tb:RDA_result} shows the comparison of NIE between BIMA and BASMU. We use the criteria of whether the 95\% credible interval includes 0 for active mediation voxel selection. In the bottom table of NIE size, the total, positive, and negative effects are separately reported for each method. The posterior mean of NDE under BIMA is 0.247. The NDE under BASMU is $0.183$. The NIE over all locations is 0.073 for BIMA, and 0.103 for BASMU. To check the model fitting, the $R^2$ for the BIMA outcome model \eqref{eq:reduced}  is 0.41, and the $R^2$ for the BASMU outcome model \eqref{eq:full} is 0.42. Figure \ref{fig:RDA_TIE} provides a visual illustration of the selected active mediation voxels. Web  Figure \ref{fig:RDA_scatter} gives a scatter plot of each estimated $\cE(s_j)$ between BIMA and BASMU.

Based on Table \ref{tb:RDA_result}, BASMU tends to select more active mediation voxels in all top four regions compared to BIMA. Especially in region 34 (right middle cingulum), an area for integrating memory information, BASMU selects three times more mediation voxels, although the scalar NIE for this region remains unchanged. In terms of effect size, BASMU in region 57 (left postcentral), an area for episodic memory retrieval, has a larger scale in NIE. In the scatter plot in Web  Figure \ref{fig:RDA_scatter}, we can see that the large positive voxels are usually selected by both methods, whereas BASMU gives more selection on smaller positive effects. In the coronal view of the brain image NIE Figure \ref{fig:RDA_TIE}, the center of the large active areas are usually selected by both methods, whereas the edge of those large areas tends to differ between BIMA and BASMU. In summary, after accounting for the unobserved confounders, we tend to select more active mediation voxels, and the effect of parental education on children's general cognitive ability mediated through children's brain activity takes a larger proportion in the total effect after adjusting for the unobserved confounders. The NDE also tends to decrease after adjusting for unobserved confounders.

\section{Conclusion and Discussion}\label{sec:discussion}

In this work, based on the BIMA \citep{xu2023bayesian}, we propose the BASMU framework for structured mediators to account for the unobserved confounder effects. We utilize the subject-specific spatial confounding effects as the unobserved confounders and incorporate them into the outcome model. We provide rigorous proof for the theoretical analysis on the asymptotic bias of the outcome model, and the identifiability of the BASMU model. For the estimation step, due to the complexity of the BASMU model, we propose the two-stage estimation algorithm. While full Bayesian inference from joint estimation is challenging, our two-stage estimation method yields reasonably accurate point estimates for $\beta$ and NIE, as evidenced by extensive simulation results. Alternative ways of the two-stage algorithm include bootstrap, or updating $\eta_i$ and updating all other parameters until convergence and then iteratively updating $\eta_i$ in loops until full convergence. We apply BASMU in the ABCD study and find the mediation effect takes a larger proportion after adjusting for the unobserved confounders. A discussion on the limitation of BASMU can be found in the Supplementary Materials Section~\ref{supp_sec:additional_limitation}.

\section*{Acknowledgments}
Shu Yang's work is supported by the National Science Foundation grant SES2242776. Jian Kang's work is supported by the National Institute of Health grants R01DA048993, R01MH105561, and the National Science Foundation grant IIS2123777. We use ChatGPT to check grammar and improve sentence structure.

\section*{Data Availability}
The Adolescent Brain Cognitive Development (ABCD) data is publicly available at \url{https://abcdstudy.org/}. The ABCD data requires signing a data user agreement to gain access. The preprocessed data used in our main analysis is available upon request after obtaining data user agreement from the ABCD study. 
{
\bibliographystyle{biom} 
\bibliography{ref}

@article{Cinelli2020-ai,
  title={Making sense of sensitivity: Extending omitted variable bias},
  author={Cinelli, Carlos and Hazlett, Chad},
  journal={Journal of the Royal Statistical Society Series B: Statistical Methodology},
  volume={82},
  number={1},
  pages={39--67},
  year={2020},
  publisher={Oxford University Press}
}

@article{Lindquist2012-oj,
  title={Functional causal mediation analysis with an application to brain connectivity},
  author={Lindquist, Martin A},
  journal={Journal of the American Statistical Association},
  volume={107},
  number={500},
  pages={1297--1309},
  year={2012},
  publisher={Taylor \& Francis}
}

@article{guan2023spectral,
  title={Spectral adjustment for spatial confounding},
  author={Guan, Yawen and Page, Garritt L and Reich, Brian J and Ventrucci, Massimo and Yang, Shu},
  journal={Biometrika},
  volume={110},
  number={3},
  pages={699--719},
  year={2023},
  publisher={Oxford University Press}
}

@ARTICLE{Nath2023-mv,
  title   = "A machine learning based approach towards high-dimensional
             mediation analysis",
  author  = "Nath, Tanmay and Caffo, Brian and Wager, Tor and Lindquist, Martin
             A",
  journal = "Neuroimage",
  volume  =  268,
  pages   = "119843",
  month   =  mar,
  year    =  2023
}

@article{Zhang2022-yb,
  title={Interpretable sensitivity analysis for the baron-kenny approach to mediation with unmeasured confounding},
  author={Zhang, Mingrui and Ding, Peng},
  journal={arXiv:2205.08030},
  year={2022}
}

@ARTICLE{Tchetgen2012-hf,
  title={Semiparametric theory for causal mediation analysis: efficiency bounds, multiple robustness, and sensitivity analysis},
  author={Tchetgen, Eric J Tchetgen and Shpitser, Ilya},
  journal={Annals of statistics},
  volume={40},
  number={3},
  pages={1816},
  year={2012}
}

@article{Imai2010-ih,
  title={Identification, inference and sensitivity analysis for causal mediation effects},
  author={Imai, Kosuke and Keele, Luke and Yamamoto, Teppei},
  journal={Statistical Science},
  volume={25},
  number={1},
  pages={51--71},
  year={2010},
  publisher={Institute of Mathematical Statistics},
  doi={10.1214/10-STS321}
}

@article{VanderWeele2014-hj,
  title={Mediation analysis with multiple mediators},
  author={VanderWeele, Tyler and Vansteelandt, Stijn},
  journal={Epidemiologic methods},
  volume={2},
  number={1},
  pages={95--115},
  year={2014},
  publisher={De Gruyter}
}

@article{ding2016sharp,
  title={Sharp sensitivity bounds for mediation under unmeasured mediator-outcome confounding},
  author={Ding, Peng and Vanderweele, Tyler J},
  journal={Biometrika},
  volume={103},
  number={2},
  pages={483--490},
  year={2016},
  publisher={Oxford University Press}
}

@article{xu2023bayesian,
  title={Bayesian image mediation analysis},
  author={Xu, Yuliang and Johnson, Timothy D and Heitzeg, Mary and Kang, Jian},
  journal={Journal of the American Statistical Association},
  number={just-accepted},
  pages={1--25},
  year={2026},
  publisher={Taylor \& Francis}
}

@article{Baron1986-yh,
  title={The moderator--mediator variable distinction in social psychological research: Conceptual, strategic, and statistical considerations.},
  author={Baron, Reuben M and Kenny, David A},
  journal={Journal of personality and social psychology},
  volume={51},
  number={6},
  pages={1173},
  year={1986},
  publisher={American Psychological Association}
}

@article{casey2018ABCD,
  title={The adolescent brain cognitive development (ABCD) study: imaging acquisition across 21 sites},
  author={Casey, BJ and Cannonier, Tariq and Conley, May I and Cohen, Alexandra O and Barch, Deanna M and Heitzeg, Mary M and Soules, Mary E and Teslovich, Theresa and Dellarco, Danielle V and Garavan, Hugh and others},
  journal={Developmental cognitive neuroscience},
  volume={32},
  pages={43--54},
  year={2018},
  publisher={Elsevier}
}

@article{sripada2020toward,
  title={Toward a “treadmill test” for cognition: Improved prediction of general cognitive ability from the task activated brain},
  author={Sripada, Chandra and Angstadt, Mike and Rutherford, Saige and Taxali, Aman and Shedden, Kerby},
  journal={Human brain mapping},
  volume={41},
  number={12},
  pages={3186--3197},
  year={2020},
  publisher={Wiley Online Library}
}

@article{chen2022imaging,
  title={Imaging genetic based mediation analysis for human cognition},
  author={Chen, Tingan and Mandal, Abhishek and Zhu, Hongtu and Liu, Rongjie},
  journal={Frontiers in neuroscience},
  volume={16},
  pages={824069},
  year={2022},
  publisher={Frontiers}
}

@article{wang2023high,
  title={A high-dimensional mediation model for a neuroimaging mediator: Integrating clinical, neuroimaging, and neurocognitive data to mitigate late effects in pediatric cancer},
  author={Wang, Jade Xiaoqing and Li, Yimei and Reddick, Wilburn E and Conklin, Heather M and Glass, John O and Onar-Thomas, Arzu and Gajjar, Amar and Cheng, Cheng and Lu, Zhao-Hua},
  journal={Biometrics},
  volume={79},
  number={3},
  pages={2430--2443},
  year={2023},
  publisher={Oxford University Press}
}

@article{jiang2023causal,
  title={Causal mediation analysis using high-dimensional image mediator bounded in irregular domain with an application to breast cancer},
  author={Jiang, Shu and Colditz, Graham A},
  journal={Biometrics},
  volume={79},
  number={4},
  pages={3728--3738},
  year={2023},
  publisher={Oxford University Press}
}

@article{song2020cors,
  title={Bayesian hierarchical models for high-dimensional mediation analysis with coordinated selection of correlated mediators},
  author={Song, Yanyi and Zhou, Xiang and Kang, Jian and Aung, Max T and Zhang, Min and Zhao, Wei and Needham, Belinda L and Kardia, Sharon LR and Liu, Yongmei and Meeker, John D and others},
  journal={Statistics in medicine},
  volume={40},
  number={27},
  pages={6038--6056},
  year={2021},
  publisher={Wiley Online Library}
}

@article{song2021ptg,
  title={Bayesian sparse mediation analysis with targeted penalization of natural indirect effects},
  author={Song, Yanyi and Zhou, Xiang and Kang, Jian and Aung, Max T and Zhang, Min and Zhao, Wei and Needham, Belinda L and Kardia, Sharon LR and Liu, Yongmei and Meeker, John D and others},
  journal={Journal of the Royal Statistical Society Series C: Applied Statistics},
  volume={70},
  number={5},
  pages={1391--1412},
  year={2021},
  publisher={Oxford University Press}
}

@article{gao2019testing,
  title={Testing mediation effects in high-dimensional epigenetic studies},
  author={Gao, Yuzhao and Yang, Haitao and Fang, Ruiling and Zhang, Yanbo and Goode, Ellen L and Cui, Yuehua},
  journal={Frontiers in genetics},
  volume={10},
  pages={1195},
  year={2019},
  publisher={Frontiers Media SA}
}

@Manual{bama,
    title = {bama: High Dimensional Bayesian Mediation Analysis},
    author = {Alexander Rix and Mike Kleinsasser and Yanyi Song},
    year = {2023},
    note = {R package version 1.3.0},
    url = {https://CRAN.R-project.org/package=bama},
  }

@Article{hdmed,
    title = {Methods for Mediation Analysis with High-Dimensional DNA Methylation Data: Possible Choices and Comparison},
    author = {Dylan Clark-Boucher and Xiang Zhou and Jiacong Du and Yongmei Liu and Belinda L Needham and Jennifer A Smith and Bhramar Mukherjee},
    year = {2023},
    journal = {PLOS Genetics},
    volume = {19},
    issue = {11},
    pages = {1-26},
    url = {https://doi.org/10.1371/journal.pgen.1011022},
    doi = {10.1371/journal.pgen.1011022},
  }

@article{zeng2021statistical,
  title={Statistical methods for mediation analysis in the era of high-throughput genomics: current successes and future challenges},
  author={Zeng, Ping and Shao, Zhonghe and Zhou, Xiang},
  journal={Computational and structural biotechnology journal},
  volume={19},
  pages={3209--3224},
  year={2021},
  publisher={Elsevier}
}

@ARTICLE{Dorie2016-om,
  title    = "A flexible, interpretable framework for assessing sensitivity to
              unmeasured confounding",
  author   = "Dorie, Vincent and Harada, Masataka and Carnegie, Nicole Bohme
              and Hill, Jennifer",
  journal  = "Stat. Med.",
  volume   =  35,
  number   =  20,
  pages    = "3453--3470",
  month    =  sep,
  year     =  2016,
  language = "en"
}

@article{rudolph2024using,
  title={Using instrumental variables to address unmeasured confounding in causal mediation analysis},
  author={Rudolph, Kara E and Williams, Nicholas and D{\'\i}az, Iv{\'a}n},
  journal={Biometrics},
  volume={80},
  number={1},
  pages={ujad037},
  year={2024},
  publisher={Oxford University Press}
}

@article{angrist1996identification,
  title={Identification of causal effects using instrumental variables},
  author={Angrist, Joshua D and Imbens, Guido W and Rubin, Donald B},
  journal={Journal of the American statistical Association},
  volume={91},
  number={434},
  pages={444--455},
  year={1996},
  publisher={Taylor \& Francis}
}

@article{lipsitch2010negative,
  title={Negative controls: a tool for detecting confounding and bias in observational studies},
  author={Lipsitch, Marc and Tchetgen, Eric Tchetgen and Cohen, Ted},
  journal={Epidemiology},
  volume={21},
  number={3},
  pages={383--388},
  year={2010},
  publisher={LWW}
}

@article{miao2024confounding,
  title={A confounding bridge approach for double negative control inference on causal effects},
  author={Miao, Wang and Shi, Xu and Li, Yilin and Tchetgen Tchetgen, Eric J},
  journal={Statistical Theory and Related Fields},
  volume={8},
  number={4},
  pages={262--273},
  year={2024},
  publisher={Taylor \& Francis}
}

@article{vanderweele2014effect,
  title={Effect decomposition in the presence of an exposure-induced mediator-outcome confounder},
  author={VanderWeele, Tyler J and Vansteelandt, Stijn and Robins, James M},
  journal={Epidemiology},
  volume={25},
  number={2},
  pages={300--306},
  year={2014},
  publisher={LWW}
}
}

\newpage

\section*{Supplementary Materials}
 The BASMU package (\url{https://github.com/yuliangxu/BASMU}
) includes installation instructions, reproducible simulation examples, and visualization code.  

\global\long\def\thefigure{S\arabic{figure}}
\setcounter{figure}{0}
\global\long\def\thetable{S\arabic{table}}
\setcounter{table}{0}
\global\long\def\thesection{S\arabic{section}}
\setcounter{section}{0}
\renewcommand{\theequation}{S.\arabic{equation}}
\setcounter{equation}{0}

\section{Table~\ref{tb:notation} of Notations for Quick Reference }

\begin{table}[ht!]
\centering
\caption{Table of Notations}
\label{tb:notation}
\begin{tabular}{|c|l|}
\hline
\textbf{Notation} & \textbf{Description} \\ 
\(X\) & Exposure or treatment variable \\ 
\(M\) & Mediator variable with spatial structure, such as brain imaging data \\ 
\(Y\) & Outcome variable \\ 
\(C\) & Observed confounders \\ 
\(\eta_i\) & subject-specific spatial confounding effects \\
& representing unobserved confounding factors \\ 
\(\nu\) & Effect of unobserved confounders on the outcome \\ 
\(\alpha(s_j)\) & Impact of the treatment \(X\) on the mediator at location \(s_j\) \\ 
\(\xi_k(s_j)\) & Functional coefficient for the observed confounders \(C\) on \(M\) at location \(s_j\) \\ 
\(\beta(s_j)\) & Effect of the image mediator \(M\) on the outcome \(Y\) at location \(s_j\) \\ 
\(\gamma\) & Direct effect of \(X\) on \(Y\) \\ 
\(\zeta_k\) & Coefficient for the \(k\)-th observed confounder \(C_k\) on \(Y\) \\ 
\(\epsilon_M(s_j)\) & Spatially-independent noise term in the mediator model at location \(s_j\) \\ 
\(\epsilon_Y\) & Error term in the outcome model \\
\(\lambda(\Delta s_j)\) & Lebesgue measure on the pixel partition \( \Delta s_j \) \\
\(\text{GP}(0, \kappa)\) & Gaussian Process with mean function 0 and covariance function \(\kappa\) \\
\(\theta_{\eta, l}\) & Basis coefficients for the Gaussian Process prior \\
& for subject-specific spatial confounding effects \( \eta \) \\ 
\(\psi_l(s)\) & Basis functions for the Gaussian Process prior \\
\(\theta_{all}\) & Collection of all parameters in models \\ 
\(U_0\) & True unobserved confounder term \\
\(\hat{U}\) & Estimator of the unobserved confounder term \\ 
\(\text{MSE}\) & Mean Squared Error \\ 
\(\text{NIE}\) & Natural Indirect Effect \\ 
\(\text{NDE}\) & Natural Direct Effect \\ 
\(\sigma_M\) & Variance of the noise term in the mediator model \\ 
\(\sigma_Y\) & Variance of the noise term in the outcome model \\ 
\(\sigma_\eta\) & Variance of the Gaussian Process prior for \(\eta\) \\ 
\(\sigma_\nu\) & Variance of the Gaussian Process prior for \(\nu\) \\ 
\(\delta(s)\) & Selection variable for the prior on \(\nu\) \\ 
\(\cE(s)\) & Spatially-varying NIE \\ 
\(\cE\) & Scalar-valued NIE \\ \hline
\end{tabular}
\end{table}

\section{Proof of Proposition \ref{prop:identifiable}}
\begin{proof}
Denote $\theta = \sbr{\alpha,\beta,\xi,\zeta,\eta, \nu,\sigma_M, \sigma_Y}$ as the collection of all parameters.
First of all, based on model \eqref{eq:mediator} and the fact that for the intensity measure of $M_i(\Delta s_j)$ over a small voxel partition $\Delta s_j$ follows Gaussian distribution $M_i(\Delta s_j) = \bE\br{M_i(s_j)}\leb(\Delta s_j) + \epsilon_{M,i}(\Delta s_j)$, the mean and variance function of $M_i(\Delta s_j)$ can be uniquely identified, hence $\sigma_M$ and the mean function $\bE\br{M_i(s_j)\mid\theta,X_i,C_i}$ are both uniquely identifiable. We denote the mean function $\bE\br{M_i(s_j)\mid\theta,X_i,C_i}$ as 
\begin{align*}
    A_{i,j}(\alpha,\xi,\eta_i) &= X_i\alpha(s_j) + \sum_{k=1}^q C_{i,k}\xi_{k}(s_j) + \eta_i(s_j)
\end{align*}

Similarly, $\sigma_Y$ is also uniquely identifiable. In the derivation below, for simplicity we denote 
\begin{align*}
    B_i(\gamma, \zeta,\nu,\eta_i)&= X_i\gamma + \sum_{k=1}^qC_{i,k}\zeta_k + \sum_{j=1}^p\nu(s_j)\eta_i(s_j)\leb(\Delta s_j)
\end{align*}

Let $\bfM_i = \br{M_i(\Delta s_j)}_{j=1}^p$.
Conditional on the covariates $X_i, C_{i,k}$, 
the joint distribution of $Y_i, \bfM_i$ can be expressed as 
\begin{align*}
    &\pi\sbr{Y_i, \bfM_i \mid X_i, \br{C_i}_{k=1}^m, \theta} =  
    \pi\sbr{Y_i \mid \bfM_i ,X_i, \br{C_i}_{k=1}^m, \theta} \prod_{j=1}^p\pi\sbr{M_i(\Delta s_j)\mid X_i, \br{C_i}_{k=1}^m, \theta}\\
    =&\frac{1}{\sqrt{2\pi\sigma_Y^2}} \exp\br{-\frac{1}{2\sigma_Y^2}\sbr{Y_i - \sum_{j=1}^p M_i(\Delta s_j)\beta(s_j) - B_i(\gamma, \zeta,\nu,\eta_i)}^2} \\
    &\times \prod_{j=1}^p\sbr{\frac{1}{\sqrt{2\pi\sigma_M^2\leb(\Delta s_j)}}} \times\exp\br{-\frac{1}{2\sigma_M^2}\sum_{j=1}^p\frac{1}{\leb(\Delta s_j)}\sbr{M_i(\Delta s_j) - A_{i,j}(\alpha,\xi,\eta_i)}^2}
\end{align*}
Suppose that $Y_i$ takes value $y_i$ and $M_i(\Delta s_j)$ takes value $m_{i,j}$ in the joint density function, note that $y_i$ and $m_{i,j}$ can be any real values. We can write the log joint density function over the $i$-th individual $i=1,\dots,n$ as 
\begin{align*}
    &\sum_{i=1}^n \br{\log \pi(y_i, \br{m_{i,j}}_{j=1}^p \mid X_i, \br{C_{i,k}}_{k=1}^q, \theta) } \\
    \propto 
    &\sum_{i=1}^n \Bigg\{ -\frac{1}{2\sigma_Y^2}y_i^2 - \frac{1}{2\sigma_Y^2}\sbr{\sum_{j=1}^pm_{i,j}\beta(s_j)}^2 - \frac{1}{2\sigma_Y^2}B_i^2(\gamma, \zeta,\nu,\eta_i) \\
    &+\frac{1}{\sigma_Y^2} y_i\sbr{\sum_{j=1}^p m_{i,j}\beta(s_j)} + \frac{1}{\sigma_Y^2} B_i(\gamma, \zeta,\nu,\eta_i) y_i + \frac{1}{\sigma_Y^2} \sbr{\sum_{j=1}^pm_{i,j}\beta(s_j)} B_i(\gamma, \zeta,\nu,\eta_i) \\
    &-\frac{p}{2\sigma_M^2}\sum_{j=1}^p\sbr{m_{i,j}}^2 - \frac{p}{2\sigma_M^2} A_{i,j}^2(\alpha,\xi,\eta) + \frac{p}{2\sigma_M^2}\sum_{j=1}^p m_{i,j} A_{i,j}(\alpha,\xi,\eta_i) \Bigg\}
\end{align*}
This is a polynomial of $y_i, m_{i,1}, \dots,m_{i,p}$, and we only need to match the coefficients of the first-order, second-order, and interaction terms to identify the unique coefficients. Hence $\sigma_Y^2$ can be uniquely determined by the quadratic term $\sum_{i=1}^n y_i^2$, similarly $\br{\beta(s_j)}_{j=1}^p$ uniquely determined by the interaction terms $\br{\sum_{i=1}^n y_i m_{i,j}}_{j=1}^p$, and $B_i(\gamma, \zeta,\nu,\eta_i)$ uniquely determined by the first-order term $y_i$. Given that $\br{\beta(s_j)}_{j=1}^p$ and $\sigma_Y^2$ are uniquely identifiable, we can also uniquely determine $\sigma_M^2$ from the coefficient of $\sum_{i=1}^n\sum_{j=1}^p m_{i,j}^2$. Given the identified $\br{\beta(s_j)}_{j=1}^p$,  $\sigma_Y^2$ and $\sigma_M^2$, $A_{i,j}(\alpha,\xi,\eta_i)$ is also identified from the coefficient of the first-order $m_{i,j}$.

Now we have shown the identifiability of $\sigma_Y^2,\sigma_M^2,\br{\beta(s_j)}_{j=1}^p$, $A_{i,j}(\alpha,\xi,\eta_i)$ and $ B_i(\gamma, \zeta,\nu,\eta_i)$.

Next, we need to show that the rest of the parameters $\sbr{\alpha,\xi,\zeta,\eta, \nu}$ can also be uniquely identified from $A_{i,j}(\alpha,\xi,\eta_i)$ and $ B_i(\gamma, \zeta,\nu,\eta_i)$. Given Assumption \ref{asm:iden}.1-2, the identifiability of $\alpha,\xi,\eta$ in $A_{i,j}(\alpha,\xi,\eta_i)$ directly follows from Proposition 1 in \cite{xu2023bayesian}.

To show the identifiability of $\gamma, \zeta,\nu$ in $ B_i(\gamma, \zeta,\nu,\eta_i)$, note that given $B_i(\gamma, \zeta,\nu,\eta_i)$ and $\eta_i$ are identifiable for $i=1,\dots,n$, comparing $B_i(\gamma', \zeta',\nu',\eta_i) = B_i(\gamma, \zeta,\nu,\eta_i),i=1,\dots, n$ to reach the identifiability of $\gamma, \zeta,\nu$ is equivalent to solving a linear system (given that the design matrix is full rank,  Assumption \ref{asm:iden}.3 ) with $n$ equations and $p+1+q$ variables. Hence under the assumption (ii) where $\nu$ is sparse, $\nu \in \Theta^\SP$, when $n$ is large enough, the number of nonzero elements in $\nu$ will be smaller than $n-q-1$, hence $\gamma, \zeta,\nu$ are all identifiable. 

Under assumption (i), $\nu$ is spatially-correlated and can be decomposed using $L$ number of basis. Let $\nu(s) = \sum_{l=1}^L\theta_{\nu,l}\psi_l(s)$, and 
\begin{align*}
    \int_{\cS}\nu(s)\eta_i(s)\leb (\dd s)&= \int_{\cS}\sum_{l=1}^L\theta_{\nu,l}\psi_l(s)\eta_i(s)\leb (\dd s) =\sum_{l=1}^L\theta_{\nu,l}\theta_{\eta_i,l}
\end{align*}
Hence 
\[B_i(\gamma, \zeta,\nu,\eta_i) = X_i\gamma + \sum_{k=1}^qC_{i,k}\zeta_k + \sum_{l=1}^L\theta_{\nu,l}\theta_{\eta_i,l}\]

Based on Assumption \ref{asm:iden}.3, with the design matrix $B = \sbr{\bfX, \bfC_1,\dots, \bfC_q, \bftheta_{\eta,1},\dots ,\bftheta_{\eta,L} }\in \R^{n\times \sbr{1+L+q}}$. And $B_i(\gamma, \zeta,\nu,\eta_i) - B_i(\gamma', \zeta',\nu',\eta_i)=0$ for $i=1,\dots,n$ can be written as
\begin{align*}
    B\cdot\br{\sbr{\gamma, \zeta,\theta_{\nu,1},\dots, \theta_{\nu,L}}^\rT-\sbr{\gamma', \zeta',\theta'_{\nu,1},\dots, \theta'_{\nu,L}}^\rT} =\bf0
\end{align*}
By  Assumption \ref{asm:iden}.3, $\det(B)>0$, hence $\sbr{\gamma, \zeta,\theta_{\nu,1},\dots, \theta_{\nu,L}} = \sbr{\gamma', \zeta',\theta'_{\nu,1},\dots, \theta'_{\nu,L}}$, therefore $\gamma, \zeta,\nu$ are also identifiable. 
Similarly, if $\nu\in\Theta^\SP$, the design matrix becomes $$\br{\bfX, \bfC_1,\dots, \bfC_q, \br{\bfeta(s_k)}_{s_k\in\cS_m}}\in \R^{n\times (m+1+q)}$$ where $\cS_m=\br{s: \nu(s) \neq 0}$, and is also full rank by Assumption \ref{asm:iden}.
\end{proof}

\section{Proof of Theorem \ref{prop:bias}}
\begin{proof}

Throughout this proof, we use the notation $o_p(1)$ as follows: if $X_n=o_p(1)$, $X_n \overset{p}{\to}0$ as $n\to\infty$.

Using the decomposition on $\beta(s) =  \sum_{l=1}^L\theta_{\theta_\beta,l} \psi_l(s)$, $\tilde M_{i,l} = \int_{S}M_i(s)\psi_l(s)\leb(\dd s)$, the full outcome model can be decomposed as 
\begin{align*}
    Y_i = \sum_{l=1}^L\theta_{\beta,l}\Tilde{M}_{i,l} + \gamma X_i + \sum_{k=1}^q\zeta_k C_{i,k} + \sum_{j=1}^p\eta_i(s_j)\nu(s_j) + \epsilon_{Y,i}
\end{align*}
where $\epsilon_{Y,i}\overset{\iid}{\sim}N(0,\sigma_Y^2)$.

With the prior specification $\theta_{\beta,l}\overset{\ind}{\sim} N(0,\sigma_\beta^2\lambda_l)$, denote diagonal matrix $D\in \R^{L\times L}$, where $(D)_{l,l'}=\lambda_l I(l=l')$.
Denote $\tilde{M}_{i} = \sbr{\tilde{M}_{i,l},\dots,\tilde{M}_{i,L}}^\rT\in \R^{L}$.
The posterior mean of $\theta_\beta$ is 
\begin{align*}
    \Var\br{\theta_\beta|\sim}&= \sbr{\frac{1}{\sigma_\beta^2}D^{-1} + \frac{1}{\sigma_Y^2}\sum_{i=1}^n \tilde{M}_i \tilde{M}_i^\rT}^{-1}\\
    \bE\br{\theta_\beta|\sim} &= \Var\br{\theta_\beta|\sim}\br{\frac{1}{\sigma_Y^2}\sum_{i=1}^n \sbr{Y_i - \gamma X_i-\sum_{k=1}^q\zeta_k C_{i,k} -\sum_{j=1}^p\eta_i(s_j)\nu(s_j) }\tilde{M}_i}
\end{align*}

To simplify these two bias expressions, we denote $\tM\in \R^{n\times L}$ with each row being $\tM_i^\rT$. Let $A:= \tM^T\tM\in\R^{L\times L}$.

Denote the point estimator $\hat\theta_\beta^F = \bE\br{\theta_\beta|\sim}$ for the posterior mean under full model \eqref{eq:full}. Denote $\theta^0$ as the true parameters.
Conditional on the estimator $\hat\eta$ and $\hat\nu$, the bias of $\hat\theta_\beta^F $ can be written as 
\begin{align*}
    \bias(\hat\theta_\beta^F ) &= \bE\sbr{\hat\theta_\beta^F - \theta_\beta^0}\\
    &=\bE_{Y|\br{X,C,M}}\sbr{\hat\theta_\beta^F} - \theta_\beta^0\\
    &= \Var\br{\theta_\beta|\sim} \frac{1}{\sigma_Y^2}\sum_{i=1}^n \br{\sbr{\theta_\beta^0}^\rT\tM_i + \sbr{\nu^0}^\rT\eta_i^0 - \sbr{\hat\nu}^\rT\hat\eta_i} \tM_i - \theta_\beta^0
\end{align*}
Similarly, if we denote the point estimator using BIMA model as $\hat\theta_\beta^B$,
\begin{align*}
    \bias(\hat\theta_\beta^B ) &=\Var\br{\theta_\beta|\sim} \frac{1}{\sigma_Y^2}\sum_{i=1}^n \br{\sbr{\theta_\beta^0}^\rT\tM_i + \sbr{\nu^0}^\rT\eta_i^0 } \tM_i - \theta_\beta^0
\end{align*}

Recall the notation $\tM\in \R^{n\times L}$ and $A:= \tM^T\tM\in\R^{L\times L}$, we can simplify $\bias(\hat\theta_\beta^F )$. By the singular value assumption of $A$ in Theorem \ref{prop:bias}, with probability $1-\exp\br{-c_0 n}$, $A$ is full rank. Conditioning on the event that $A$ is full rank hence invertible, denote $\bfeta^0 \in \R^{n\times p}$ with the $i$-th row being $\eta_i^\rT$.
\begin{align*}
    \bias(\hat\theta_\beta^B ) &= \mbr{\frac{\sigma_Y^2}{\sigma_\beta^2}D^{-1} +  A}^{-1} \mbr{A\theta_\beta^0 + \sbr{\tM}^\rT \sbr{\bfeta^0}^\rT\nu^0} - \theta_\beta^0 \\
    &= \mbr{\frac{\sigma_Y^2}{\sigma_\beta^2}D^{-1}A^{-1} +  I_L}^{-1} \mbr{\theta_\beta^0 + A^{-1}\sbr{\tM}^\rT \sbr{\bfeta^0}^\rT\nu^0} - \theta_\beta^0 \\
    &\overset{(*)}{=} \mbr{I_L - \sbr{\tau^2DA + I_L}^{-1}} \mbr{\theta_\beta^0 + A^{-1}\sbr{\tM}^\rT \sbr{\bfeta^0}^\rT\nu^0} - \theta_\beta^0 \\
    &=A^{-1}\sbr{\tM}^\rT \sbr{\bfeta^0}^\rT\nu^0 - \sbr{\tau^2DA + I_L}^{-1}\mbr{\theta_\beta^0 - A^{-1}\sbr{\tM}^\rT \sbr{\bfeta^0}^\rT\nu^0}
\end{align*}
Note that $(*)$ uses the Identity $(I+A)^{-1} = I - A(I+A)^{-1} = I - (A^{-1}+I)^{-1}$, and the notation $\tau^{-2} = \frac{\sigma_Y^2}{\sigma_\beta^2}$.

\textbf{Proof of Part (i)}

Now we can see that if $\sbr{\bfeta^0}^\rT\nu^0 = \bfzero$, i.e. when the unmeasured confounder effect does not exist, the bias of $\hat\theta_\beta^F$ becomes
\begin{align*}
    \bias(\hat\theta_\beta^B )  &= - \sbr{\tau^2DA + I_L}^{-1}\theta_\beta^0
\end{align*}
The range of $\bias(\hat\theta_\beta^B )$ is controlled by the smallest and  largest eigen-values of $(\tau^2DA + I_L)^{-1}$, scaled up to a rotation of $\theta_\beta^0$. Note that $\sigma_{\min}(D)\sigma_{\min}(A)\leq \sigma_{\min}(DA)\leq \sigma_{\max}(DA)\leq \sigma_{\max}(D)\sigma_{\max}(A)$. With the assumption $\lambda_L > c_{\lambda} n^{-1+a_\lambda}$, use $h\gtrsim g$ to denote the inequality $h > c_g g$ up to a positive constant $c_g$ that does not contain any rate of $n$. We can see that $\sigma_{\min}(D)\sigma_{\min}(A) \gtrsim n^{a_\lambda} \to \infty$ as $n\to \infty$, and  $\sigma_{\max}(D)\sigma_{\max}(A)\lesssim n \to\infty$, hence $\bias(\hat\theta_\beta^B) \to \bfzero$ as $n\to\infty$.

\textbf{Proof of Part (ii)}

Similarly, when $\sbr{\bfeta^0}^\rT\nu^0 \neq \bfzero$, 
\begin{align*}
    \bias(\hat\theta_\beta^B ) &= \mbr{\tau^{-2}D^{-1} + A}^{-1} \sbr{\tM}^\rT \sbr{\bfeta^0}^\rT\nu^0  - \sbr{\tau^2DA + I_L}^{-1}\theta_\beta^0 \\
    \bias(\hat\theta_\beta^F ) &= \mbr{\tau^{-2}D^{-1} + A}^{-1} \sbr{\tM}^\rT \br{\sbr{\bfeta^0}^\rT\nu^0 -\sbr{\hat\bfeta}^\rT\hat\nu } - \sbr{\tau^2DA + I_L}^{-1}\theta_\beta^0 
\end{align*}
As we've shown that $\sbr{\tau^2DA + I_L}^{-1}\theta_\beta^0 = o_p(1)$, we focus on the first term in both bias expressions.

For the image mediator $M_i$, the mediator model \eqref{eq:mediator} assumes that $M_i(s_j) = \bE\br{M_i(s_j)} + \sigma_M Z_{i,j}$, where $Z_{i,j}$ are the independent standard normal variables. Under the orthonormal decomposition, $\tM_{i,l} = \mu_{i,l} + \sigma_M Z_{i,l}$ where the $Z_{i,l}$ are still independent standard normal under orthonormal transformation, and $\mu_{i,l}$ is a constant mean term that determines mean structure of $\tM_{i,l}$.
Hence we can write $\tM = \bfmu + \sigma_M\bfZ \in \R^{n\times L}$.
\begin{align*}
    &\mbr{\tau^{-2}D^{-1} + A}^{-1} \sbr{\tM}^\rT \sbr{\bfeta^0}^\rT\nu^0  =\\
    &\br{\frac{1}{n}\mbr{\tau^{-2}D^{-1} + \sbr{\bfmu + \sigma_M\bfZ}^\rT \sbr{\bfmu + \sigma_M\bfZ} } }^{-1} \br{\frac{1}{n} \sbr{\bfmu + \sigma_M\bfZ}^\rT\sbr{\bfeta^0}^\rT\nu^0 }
\end{align*}

The denominator can be broken down into 4 parts,
\begin{align*}
    &\frac{1}{n}\mbr{\tau^{-2}D^{-1} + \sbr{\bfmu + \sigma_M\bfZ}^\rT \sbr{\bfmu + \sigma_M\bfZ} }\\
    =&\frac{1}{n} \tau^{-2}D^{-1} + \frac{1}{n}\sum_{i=1}^n\mu_i \mu_i^\rT + \frac{\sigma_M^2}{n}\sum_{i=1}^n Z_iZ_i^\rT + \frac{\sigma_M}{n}\sum_{i=1}^n\sbr{\mu_iZ_i^\rT + Z_i\mu_i^\rT}
\end{align*}

The first term $\frac{1}{n} \tau^{-2}D^{-1} = o_p(1)$ since $1/n/\lambda_L \lesssim n^{-a_\lambda} \to 0$. 

By Assumption \ref{asm:M_mean}, the second term has a constant limit, where $\mu_{i,l} = \int_{\cS} \br{E_i(s)\psi_l(s)} \lambda(\dd s)$. We denote $H=\lim_{n\to \infty} \frac{1}{n}\sum_{i=1}^n\mu_i \mu_i^\rT \in \R^{L\times L}$, where the $(l,l')$-th element in $H$ is $H_{l,l'}$, a finite constant introduced in Assumption \ref{asm:M_mean}.

The third term has the limit $\frac{\sigma_M^2}{n}\sum_{i=1}^n Z_iZ_i^\rT \overset{p}{\to}\sigma_M^2I_L$ due to the i.i.d. normality of $\bfZ$. The last term is also $o_p(1)$, because the $(l,l')$-th term is a normal variable with mean 0 and variance $\frac{1}{n^2}\sum_{i=1}^n \sbr{\mu_{i,l}^2 + \mu_{i,l'}^2}$ for $l\neq l'$, and $\frac{4}{n^2}\sum_{i=1}^n \mu_{i,l}^2 $ for $l=l'$. Because $(H)_{l,l} = \lim_{n\to\infty}\frac{1}{n}\sum_{i=1}^n \sbr{\mu_{i,l}^2}$ is a constant, $(H)_{l,l}/n\to 0$.

Hence the denominator is equivalent to $o_p(1) + H + \sigma_M^2I_L$.

To analyze the numerator and simplify the notations, we use $U^0 := \sbr{\bfeta^0}^\rT\nu^0 \in \R^{n}$, similarly, $\hat U := \sbr{\hat\bfeta}^\rT\hat\nu$ to denote the unmeasured confounder term. The numerator can be expressed as 
\begin{align*}
    \frac{1}{n} \sbr{\bfmu + \sigma _M\bfZ}^\rT\sbr{\bfeta^0}^\rT\nu^0  &=
    \sbr{\frac{1}{n} \sum_{i=1}^n \br{\mu_{i,l} + \sigma_M Z_{i,l}} U_i^0}_{l=1}^L
\end{align*}

Note that $\frac{\sigma_M}{n}\sum_{i=1}^n Z_{i,l} U_i^0\overset{p}{\to}\bf0$ for all $l$ since $\sum_{i=1}^n\sbr{ U_i^0}^2/n$ is finite (assumption made in part (ii) of Theorem \ref{prop:bias}). 

With the Assumption \ref{asm:M_mean} that $h^0 = \lim_{n\to \infty}\sum_{i=1}^n \mu_i U_i^0$ is a finite vector in $\R^L$, we can draw the conclusion that $ \bias(\hat\theta_\beta^B ) \overset{p}{\to} (H+\sigma_M^2 I_L)^{-1}h^0$.

Similarly, if we define $\hat h = \lim_{n\to \infty}\sum_{i=1}^n \mu_i \hat U_i$, the bias of $\theta_\beta$  under the joint model \eqref{eq:full} becomes $ \bias(\hat\theta_\beta^F ) \overset{p}{\to} (H+\sigma_M^2 I_L)^{-1}\sbr{h^0-\hat h}$.

\end{proof}

\section{Two-stage Algorithm: update of $\nu$ in the outcome model}

Let $\hat\bfeta\in \R^{p\times n}$ be the estimator of $\bfeta$ obtained using the mediator model. We present details for updating $\nu$ and $\delta_\nu$ in Algorithm \ref{algo:two_stage}. Updating $\nu$ requires fast linear regression using SVD on $\eta$, which can be split into two cases, one with the nonzero element in $\nu$ greater than $n$, and the other smaller than $n$.

\begin{algorithm}
\caption{Two-stage Algorithm}
\label{algo:two_stage}
\scalebox{0.8}{
    \begin{minipage}{\linewidth}
    \centering
\begin{algorithmic}
\For{Iterations $t=1,2,...$}
    \State update $\theta_\beta$ according to the posterior derivations. 
    \State \textbf{Section 1: Update $\nu$ using SVD on the design matrix.} 
    \State Let $\delta_1 = \br{s_j: \nu(s_j)\neq 0}$ and $\delta_0 = \br{s_j: \nu(s_j)= 0}$. 
    \State Let $|\delta_1|$ be the length of $\delta_1$, and $|\delta_0|$ be the length of $\delta_0$. 
    \State Denote $\hat\bfeta^\rT_1 =\sbr{\hat\bfeta^\rT}_{[:,j], j\in \delta_1} \in \R^{n\times |\delta_1|}$, and do an SVD on $\hat\bfeta^\rT_1 = UDV^\rT$. 
    \State Let $\bfY_\nu = \bfY - \gamma\bfX - \bfC^\rT\zeta - \bfM^\rT\beta\in \R^n$ be the residual without the $\bfeta^\rT\nu$ term, and let $\bfY_\nu^* = U^\rT\bfY_\nu$. 
    \If{$|\delta_1| >n$ }
        \State Let $\tau^2 = \sigma_\nu^2/\sigma_Y^2.$ $\alpha_1\sim \bfN_{|\delta_1|}(0,\sigma_\nu^2\bfI_{|\delta_1|})$, and sample $\alpha_2\sim \bfN_n(0,\sigma_Y^2\bfI_n)$. 
        \State Set $\nu^* = \alpha_1 + \tau^2 VD(1+\tau^2D^2)^{-1}(\bfY_\nu^* - DV^\rT\alpha_1 - \alpha_2)$. $\nu_{[j], j\in\delta_1  } = \nu^*$. 
    \Else ~ Sample $\nu^*\sim \bfN_{|\delta_1|}(E_1,V_1)$, where $V_1 = \sbr{\sigma_Y^{-2}D^2 + \sigma_\nu^{-2} \bfI_{|\delta_1|}}^{-1}$, $\quad E_1 = V_1\sbr{\sigma_Y^{-2} D\bfY_\nu^*}.$ Set $\nu_{[j], j\in\delta_1  } = V \nu^*$.
    \EndIf
    \State Sample $\nu^0\sim \bfN_{|\delta_0|}(0,\sigma_\nu^2\bfI_{|\delta_0|})$, and let $\nu_{[j], j\in\delta_0  } = \nu^0$.
    \State Save $\nu$ as the $t$-th sample $\nu^{(t)}$.
    \State \textbf{Section 2: Update $\delta_\nu$ sequentially.}
    \State Let $p_\delta$ be the hyper-parameter for the Bernoulli prior on $\delta_\nu$. Here we set $p_\delta=0.5$. 
    \State Compute the residual vector as $R = \bfY_\nu - \hat\bfeta^\rT (\nu*\delta_\nu)\in \R^n$.
    \For{location $j= 1,\dots, p$}
        \If{$\delta_{\nu,j} =1$}
        $R_1 = R$, $R_0 = R + \sbr{\hat\bfeta^\rT}_{j}*\nu_j$;
        \Else ~ $R_1 = R - \sbr{\hat\bfeta^\rT}_{j}*\nu_j$, $R_0 = R$; 
        \EndIf
    \State $\log l_1 = -0.5/\sigma_Y^2 * \|R_1\|_2^2$, $\log l_0 = -0.5/\sigma_Y^2 *\|R_0\|_2^2$.
    \State $p_1 = \exp^{\log l_1 - \log l_0}$, $p_1 = p_1 *p_\delta/(1-p_\delta)$, $p_0 = 1/(p_1+1)$, $p_1 = 1-p_0$.
    \State Sample $U_j\sim \text{Unif}[0,1]$, $\delta_\nu = 1$ if $U_j<p_1$ and set $R = R_1$, otherwise $\delta_\nu = 0, R=R_0$.
    \EndFor
    \State Save $\delta_\nu$ as the $t$-th sample $\delta_\nu^{(t)}$.
    \State Update the rest of parameters $\gamma, \zeta, \sigma_Y^2, \sigma_\gamma^2, \sigma_\zeta^2,\sigma_\nu^2$ using standard Gibbs Sampler.
\EndFor \\
\Return the MCMC chains of $\theta_\beta,\nu,\delta_\nu,\gamma, \zeta, \sigma_Y^2, \sigma_\gamma^2, \sigma_\zeta^2,\sigma_\nu^2$.
\end{algorithmic}
\end{minipage}
}
\end{algorithm}

\section{A brief introduction on the competing methods used in Simulation II in Section~\ref{subsec:sim_lowd}}\label{supp_sec:describe_competing_methods}
PTG (Product-Threshold-Gaussian), implemented in the R package \texttt{bama} \citep{bama}, places a joint prior on the pair of parameters $\alpha(s_j),\beta(s_j)$, but ignores the spatial correlation across different locations $s_j$. HDMA (high-dimensional mediation analysis), implemented in the R package \texttt{hdmed} \citep{hdmed}, is a frequentist approach that applies SIS (Sure Independence Screening) to first screen out the high-dimensional mediator to lower dimensions, and then uses de-biased Lasso estimator for the lower-dimensional effects.

\section{Additional Simulation and Real Data Analysis Results}

\begin{figure}[ht]
    \centering
    \includegraphics[width = 0.7\textwidth]{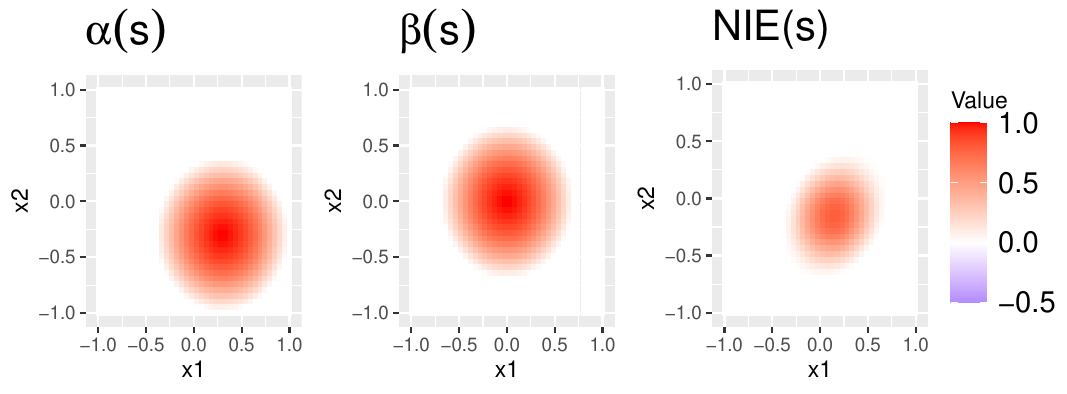}
    \caption{True signal for $\alpha(s),\beta(s)$, and spatially-varying NIE $\cE(s)$.}
    \label{fig:true_signal}
\end{figure}

\begin{table}[ht!]
\centering
\caption{Simulation result of the scalar Natural Direct Effect, averaged over 100 replications. Each column represent one method. The smallest MSE of $\cE$ is bolded in each case.}
\label{tb:sim_result_NDE}
\begin{tabular}{lcc|lcc}
\toprule
       & \textbf{BIMA}                & \textbf{BASMU}              &        & \textbf{BIMA}                        & \textbf{BASMU}                        \\
Case 1 & \multicolumn{2}{c|}{dense $\nu$}  & Case 4 & \multicolumn{2}{c}{sparse $\nu$, $n=600$} \\
Bias   & -0.44                        & -1.71                       & Bias   & -8.16                                 & -2.42                                \\
Var    & 0.04                         & 0.01                        & Var    & 0.07                                  & 0.03                                 \\
MSE    & 0.23                         & 2.94                        & MSE    & 66.71                                 & 5.88                              \\
Case 2 & \multicolumn{2}{c|}{sparse $\nu$} & Case 5 & \multicolumn{2}{c}{dense $\nu$, $\sigma_\eta=1$}    \\
Bias   & -5.14                        & -2.24                       & Bias   & -5.90                                 & -2.29                                \\
Var    & 0.06                         & 0.02                        & Var    & 0.04                                  & 0.02                                 \\
MSE    & 26.53                        & 5.05                        & MSE    & 34.85                                 & 5.25                                 \\

Case 3 & \multicolumn{2}{c|}{all 0 $\nu$}  & Case 6 & \multicolumn{2}{c}{dense $\nu$, $\sigma_M=4$}       \\
Bias   & 0.00                         & -0.26                       & Bias   & -0.51                                 & -1.70                                \\
Var    & 0.00                         & 2.23                        & Var    & 0.03                                  & 0.01                                 \\
MSE    & 0.00                         & 2.28                        & MSE    & 0.30                                  & 2.91                                
                    \\ 
\bottomrule
\end{tabular}
\end{table}

Below is a visualization of the MSE and bias of $\beta(s)$ over 100 replications, under all six simulation cases.
\begin{figure}[p]
    \centering
    \includegraphics[width = 0.7\textwidth]{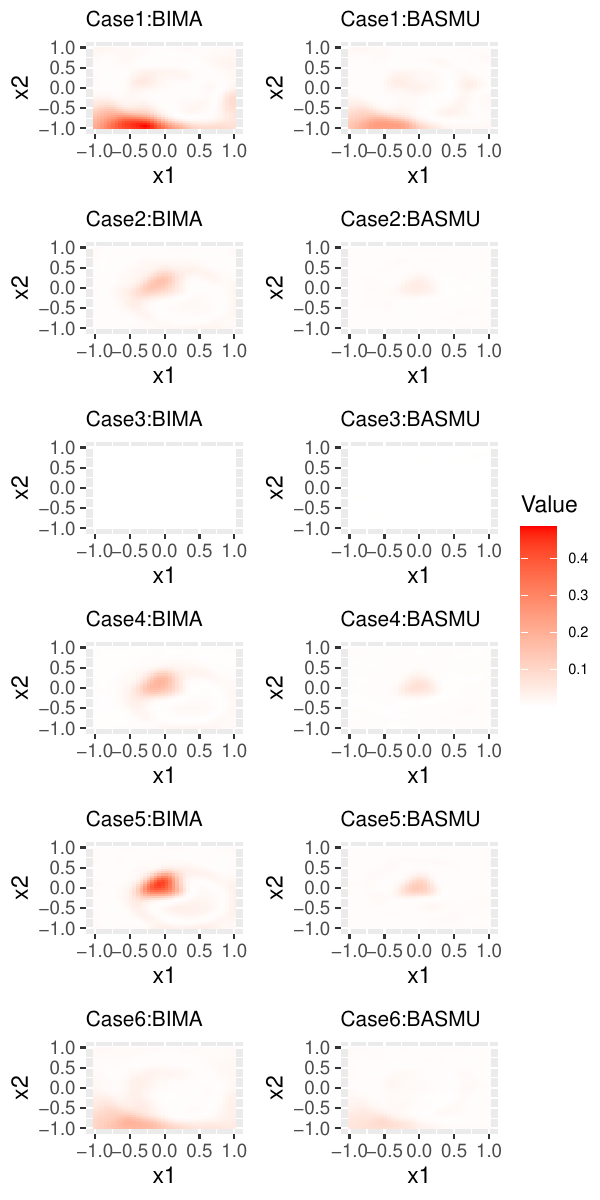}
    \caption{MSE based on 100 replications for $\beta(s)$ over different spatial locations $s$, under all simulation cases. The color bar ranges from 0 to 0.48, from white to red.}
    \label{fig:Beta_MSE}
\end{figure}

\begin{figure}[p]
    \centering
    \includegraphics[width = 0.7\textwidth]{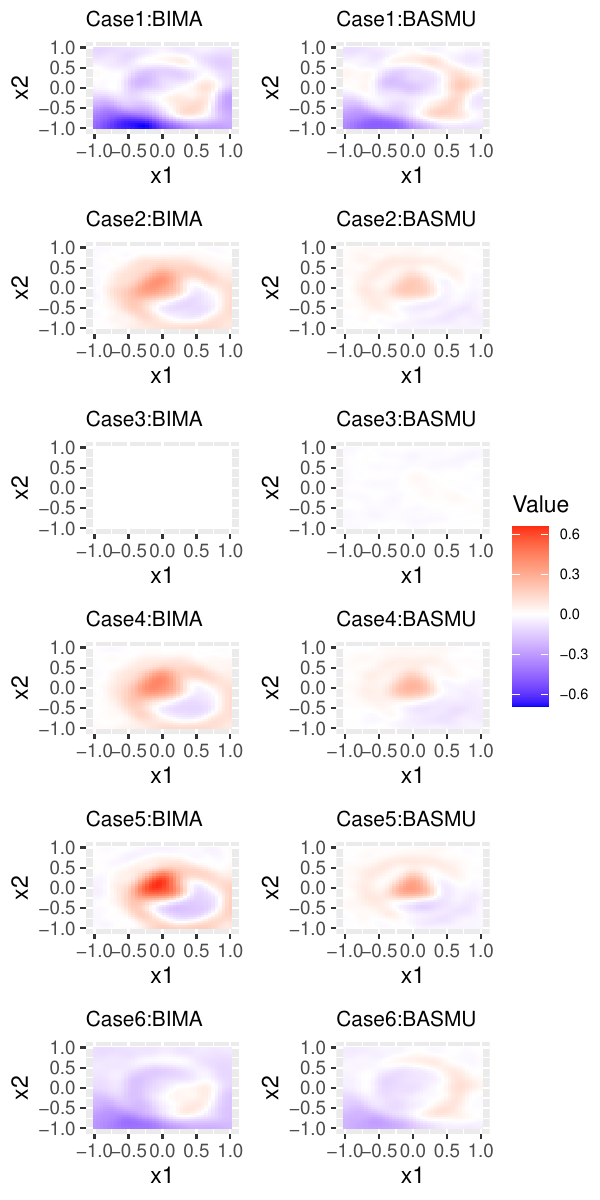}
    \caption{Bias based on 100 replications for $\beta(s)$ over different spatial locations $s$, under all simulation cases. The color bar ranges from -0.7 to 0.65, from blue (negative) to white (0) to red (positive).  }
    \label{fig:Beta_bias}
\end{figure}

\begin{figure}[p]
\centering
\begin{subfigure}[b]{\textwidth}
    \centering
    \includegraphics[width = \textwidth]{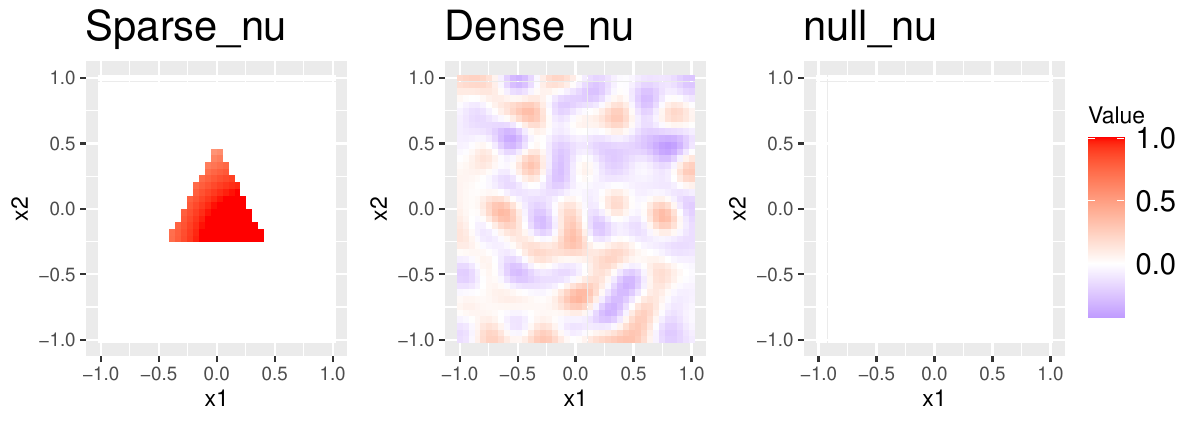}
    \caption{The true signal pattern for $\nu$, from left to right: sparse $\nu$, dense $\nu$, all 0 $\nu$.}
    \label{fig:nu_pattern}
\end{subfigure}

\begin{subfigure}[b]{\textwidth}
    \centering
    \includegraphics[width = \textwidth]{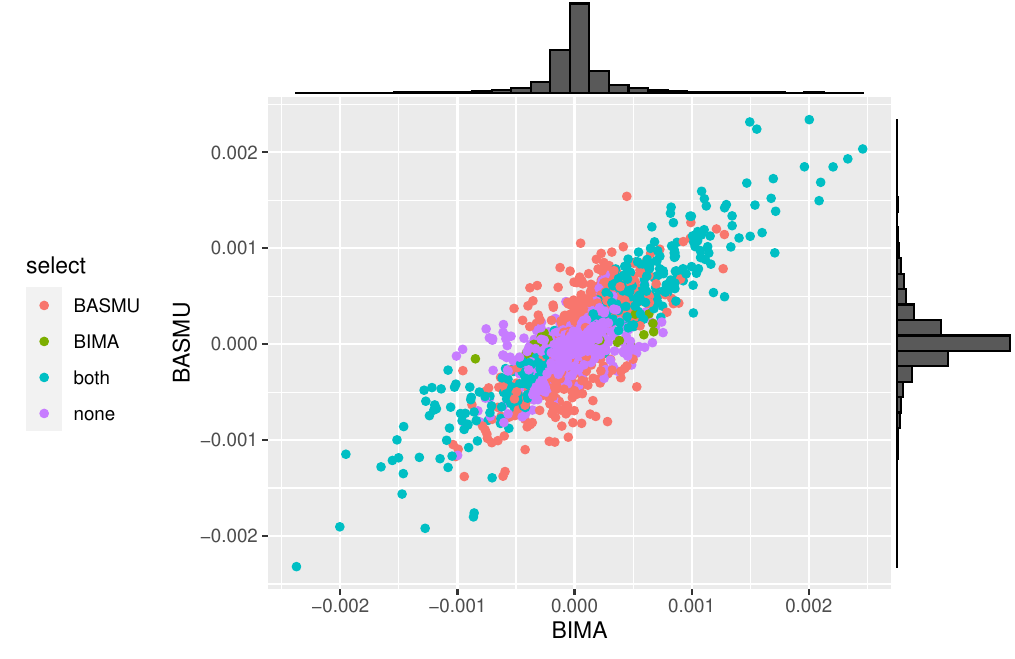}
    \caption{Scatter plot of TIE $\cE(s_j)$ comparison of BIMA and BASMU result. Each point is one voxel location. The x-axis is the value of $\cE(s_j)$ estimated by BIMA, and the y-axis is estimated by BASMU. The selection is color-coded, with the legend from top to bottom: selected only by BIMA/BASMU, selected by both methods, not selected by either method.}
    \label{fig:RDA_scatter}
\end{subfigure}
\caption{Additional simulation and real data plots.}
    
\end{figure}

\section{Empirical Validation of the Two-Stage approach}\label{sec_supp:two_stage}

The two-stage approach described in Algorithm \ref{algo:two_stage} is not a fully Bayesian approach, and this can raise problems in the uncertainty quantification using the posterior samples. The simulations in the main manuscript have demonstrated the point estimation accuracy, and this section provides additional empirical validation on the variable selection accuracy using the posterior credible interval. The key criterion of interest is to use the 95\% credible interval as the selection criterion for active spatial NIE $\cE(s)$. 

We implement three versions of the BASMU model, and compare them with BIMA to check the variable selection accuracy. The main version is Two-stage with the sparse $\nu$ prior, the same prior used in the main manuscript \eqref{eq:prior_nu}. The second version is the Two-stage algorithm with a Gaussian process prior on $\nu$, $\nu\sim \cGP(0,\sigma_\nu^2\kappa)$, where the GP kernel is the same as other functional priors in \eqref{eq:M_priors}. The third version is a fully joint update with the same GP prior $\nu\sim \cGP(0,\sigma_\nu^2\kappa)$. In the joint update, the posterior of $\eta$ is updated from the likelihoods of both \eqref{eq:full} and \eqref{eq:mediator}, and is a fully Bayesian approach. We have experimented with the joint update using the sparse $\nu$ prior, but the joint updating algorithm failed to converge due to the highly flexible $\nu$ and $\eta$ joint update, especially when they don't share the same low-dimensional structure in the case of the sparse $\nu$ prior. 

The simulation uses the sparse $\nu$ pattern as shown in Figure \ref{fig:nu_pattern}, because in practice the smoothness of $\nu$ is usually unknown, but the majority of the brain voxels should not contain significant confounding effects of the unobserved confounders. The rest of the simulation setting is as follows: $n=300$, $\sigma_M=2$, $\sigma_\eta = 0.5$. When using the posterior MCMC samples for variable selection of $\cE(s)$, if the 95\% credible interval does not contain 0, the corresponding $\cE(s)$ is deemed as having an active spatial contribution to the total NIE. We apply this criterion to all four methods: BIMA, BASMU with Two-stage and sparse $\nu$ prior, BASMU with Two-stage and GP $\nu$ prior, and BASMU with fully joint update and GP $\nu$ prior. Because the fully joint algorithm fails to converge given a sparse $\nu$ prior, we choose not to include it in the comparison. The variable selection is evaluated on the False Positive Rate (FPR), Power (or True Positive Rate), and Accuracy (the binary classification accuracy). The result based on 100 replications is reported in Table \ref{tb:sim_two_stage}.

Note that this simulation setting is even more challenging for variable selection compared to the previous point estimation. Because the true signal pattern of $\cE(s)$ as shown in Figure \ref{fig:true_signal} decays smoothly to 0. The smallest nonzero $\cE(s)$ in this case is 0.0014, which is still deemed as an active voxel even though the effect size is very small. This also fits the real data application where the fMRI signals can be very small (see Figure \ref{fig:RDA_scatter}).

\begin{table}[ht]
\centering
\caption{Simulation comparison of the variable selection results on active spatial NIE, averaged over 100 replications (standard deviations are in the brackets). FPR stands for false positive rate. Power is the true positive rate. ACC stands for accuracy, and is defined as the binary classification accuracy $\frac{1}{p}\sum_{j=1}^p I(I(\hat \cE(s_j)\neq 0) = I(\cE(s_j)\neq 0))$.}
\label{tb:sim_two_stage}
\begin{tabular}{lcccc}
\toprule
        & BIMA & \multicolumn{3}{c}{BASMU}    \\
      &  & TwoStage(Sparse) & TwoStage(GP) & Joint \\
FPR   & 0.19 (0.03) & 0.14 (0.04)             & 0.17   (0.05)      & 0.30 (0.08) \\
Power & 0.88 (0.02) & 0.87   (0.03)          & 0.86  (0.03)       & 0.94  (0.04)\\ 
ACC   & 0.82 (0.03)& 0.86    (0.03)         & 0.83  (0.04)       & 0.74 (0.06) \\ 
\bottomrule
\end{tabular}
\end{table}

As shown in Table \ref{tb:sim_two_stage}, Two-stage with sparse $\nu$ prior performs the best, with the lowest FPR and highest accuracy. Two-stage with GP prior and the joint approach both assume $\nu$ has a low-dimensional structure that can be represented by a smooth GP kernel, which is not true in this case given the true pattern of $\nu$ (see Figure \ref{fig:nu_pattern}). So the last two columns with GP priors are understandably worse than the sparse $\nu$ prior. But even with the mis-specified GP prior, the Two-stage BASMU model with GP $\nu$ prior still has better accuracy and lower FPR than BIMA, the biased model. For the joint BASMU model with GP $\nu$, although the updates of $\nu$ and $\eta$ are in the low-dimensional space (GP coefficients),  there are still too many free parameters that the joint model has the highest standard deviations and worst variable selection performance based on the 95\% CI criterion.

\begin{figure}
    \centering
        \begin{subfigure}[b]{0.45\textwidth}
            \centering
            \includegraphics[width=\textwidth]{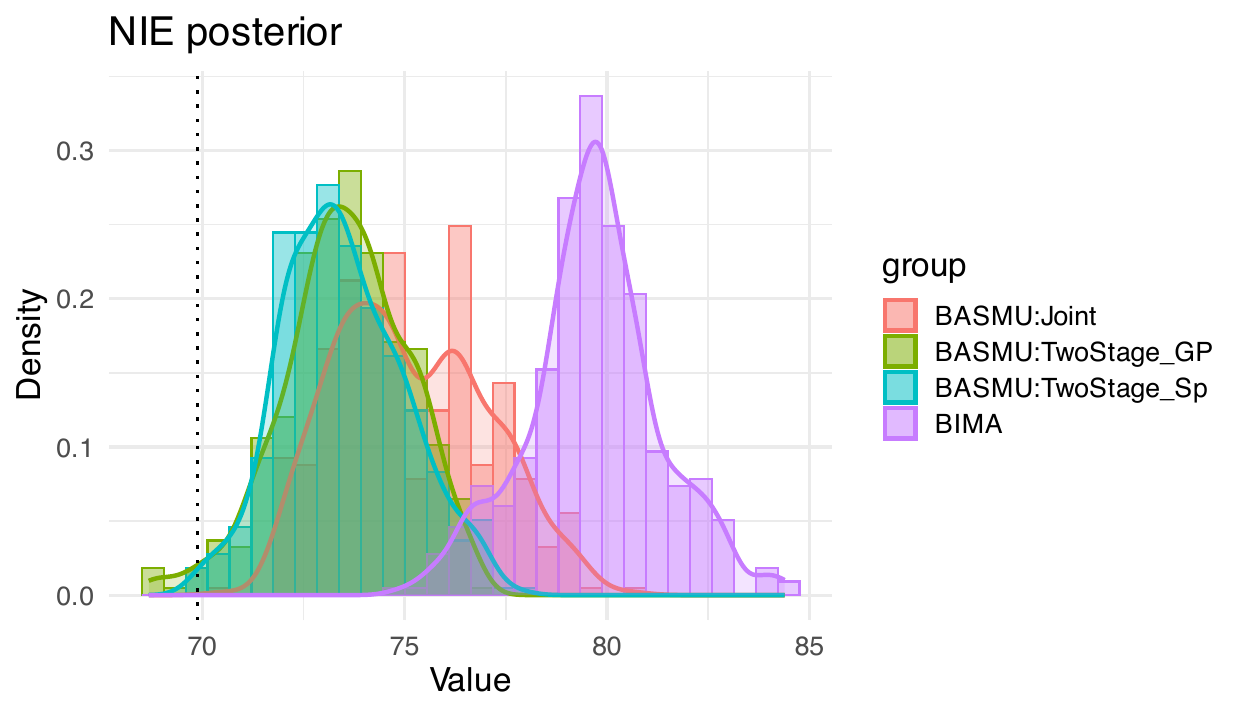}
            \caption{Posterior distribution of scalar-valued NIE based on 400 MCMC samples.}
            \label{fig:NIE_posterior}
        \end{subfigure}
  \hfill
      \begin{subfigure}[b]{0.45\textwidth}
        \centering
        \includegraphics[width=\textwidth]{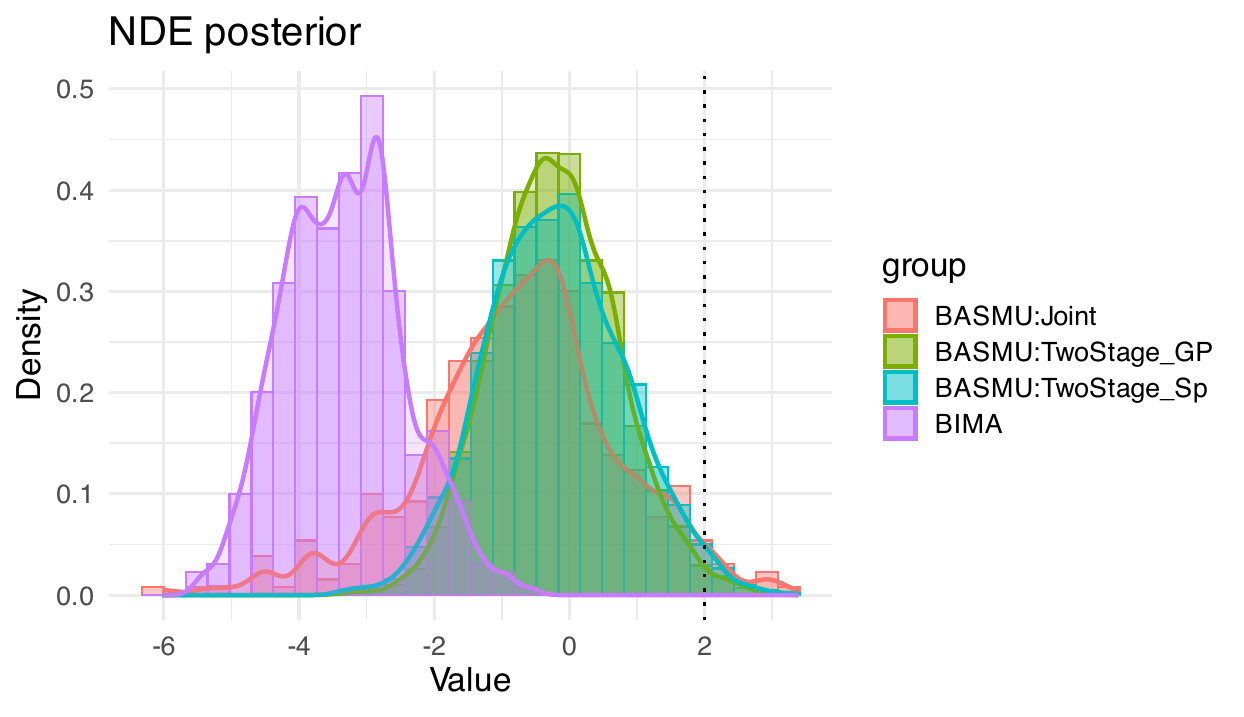}
        \caption{Posterior distribution of NDE  based on 4000 MCMC samples.}
        \label{fig:NDE_posterior}
      \end{subfigure}
    \caption{Posterior distribution of scalar-valued NIE and NDE for all methods, based on a one-time simulation. The dashed vertical line indicates the true scalar-valued NIE and NDE.}
    \label{fig:Posterior_NIE_NDE}
\end{figure}

To check and compare the posterior coverage of the Two-stage algorithms and the joint algorithm, we plot the empirical posterior distributions on the scalar-valued NIE and NDE based on MCMC samples in Figure \ref{fig:Posterior_NIE_NDE}. Note that since BIMA uses the MALA algorithm, to guarantee convergence, $\theta_\beta$ in BIMA's MCMC is thinned every 10 iterations, and we take the thinned last 400 MCMC samples based on the original 4000 iterations. Hence, the empirical NIE is based on 400 MCMC samples and the NDE is based on 4000 MCMC samples.

Figure \ref{fig:Posterior_NIE_NDE} shows that the posterior means of all three BASMU implementations are closer to the truth than the biased model BIMA. Additionally, the joint algorithm indeed has wider support compared to the two-stage algorithms. This is due to the highly flexible parameter space where $\eta$ and $\nu$ are allowed to be jointly updated. Still, the majority of the posterior mass of the joint update overlaps with that of the two-stage algorithms. This indicates that the two-stage algorithm achieves a good balance between a stable point estimation and a reasonably well-spread posterior support.

In addition, we provide the following frequentist coverage probability results on the limitation of the inference ability of using BASMU to estimate the scalar-valued NDE and total NIE. The pointwise coverage probability of a fixed parameter using a Bayesian 95\% credible interval depends on prior configuration, especially the prior variance parameters $\sigma_\gamma^2, \sigma_\beta^2, \sigma_\nu^2$. For the variance of the scalar-valued $\gamma$, we set $\sigma_\gamma$ to be a fixed value. For the high-dimensional spatial-varying parameters $\beta(s),\nu(s)$, we assign inverse-gamma priors $\sigma_\beta^2, \sigma_\nu^2\sim \text{IG}(a,b)$. \label{sent:supp_twostage}

    We have experimented with two sets of prior variance configurations,
    
    \begin{enumerate}
        \item[(i)] $\sigma_\gamma=1$, $a=b=1$ (the default): the active mediator selection result is in Table~\ref{tb:sim_two_stage}, and the frequentist pointwise coverage probability for scalar-valued NIE and NDE under 100 repeated experiments is reported in Table~\ref{tab:freq_cp_means_t} below.
        \item[(ii)] $\sigma_\gamma=2$, $a=b=0.5$: both the mediator selection results (FDR, Power, Accuracy) and the frequentist coverage probability of the scalar NIE and NDE values in Table~\ref{tab:coverage_new_prior}.
    \end{enumerate}
    Prior setting (i) is the default choice we made throughout the simulation and real data analysis. Prior setting (ii) is the more flexible prior setting, in the sense that the IG priors for $\sigma_\beta^2, \sigma_\nu^2$ place more prior mass on the tail, and the prior variance $\sigma_\gamma$ for NDE $\gamma$ is also larger.

    Based on Tables~\ref{tab:freq_cp_means_t} and \ref{tab:coverage_new_prior}, the Bayesian 95\% credible intervals in these two settings give a lower than 95\% frequentist pointwise coverage probabilities most of the time. There are two sources for this under-coverage: (a) bias, i.e., how much the posterior mean point estimate deviates from the true NIE/NDE values, and (b) variance, i.e., how concentrated the posterior mass is around the mean point estimation.  
    
    For BIMA, due to ignoring the bias induced by spatial unobserved confounders, as indicated in Figure~\ref{fig:Posterior_NIE_NDE}, the main source of low coverage is the bias shift, rather than the variance. Hence, the coverage probabilities for both NIE and NDE remain 0 for BIMA in both prior configurations in Tables~\ref{tab:freq_cp_means_t} and \ref{tab:coverage_new_prior}.

    For BASMU, when the prior settings are more flexible, comparing setting (ii) in Table~\ref{tab:coverage_new_prior} with setting (i) in Table~\ref{tab:freq_cp_means_t} and \ref{tb:sim_two_stage}, the frequentist coverage probabilities for the point estimates NIE and NDE tend to increase, and even reach 95\% for TwoStage with GP prior for $\nu$. However, because we use the Bayesian CIs for active mediator selection, a more spread-out prior mass will also lead to inflated FDR in the mediator selection, as shown in Table~\ref{tab:coverage_new_prior} compared with Table~\ref{tb:sim_two_stage}.

    \begin{table}[ht]
    \centering
    \begin{tabular}{lccc}
    \hline
     & BIMA & TwoStage\_sparse & TwoStage\_GP  \\
    \hline
    NIE (freq. CP) & 0.00 & 0.66 & 0.66 \\
    NDE (freq. CP) & 0.00 & 0.04 & 0.11 \\
    \hline
    \end{tabular}
    \caption{Frequentist Coverage Probabilities (CP) for NIE and NDE. Prior configurations (i): $\sigma_\gamma=1, a=b=1$. This is under the same simulation setting as Table~\ref{tb:sim_two_stage}.}
    \label{tab:freq_cp_means_t}
    \end{table}

    \begin{table}[ht]
    \centering
    \begin{tabular}{lccc}
    \hline
    \textbf{Metric} & \textbf{BIMA} & \textbf{TwoStage\_sparse} & \textbf{TwoStage\_GP}\\
    \hline
    FDR       & 0.46 & 0.40 & 0.45  \\
    Power      & 0.92 & 0.89 & 0.87  \\
    ACC       & 0.84 & 0.87 & 0.84  \\
    NIE (freq. CP)   & 0.00 & 0.80 & 0.79  \\
    NDE (freq. CP)   & 0.00 & 0.86 & 0.95 \\
    \hline
    \end{tabular}
    \caption{Averaged active mediator selection metrics (FDR, Power, Accuracy) and frequentist coverage probabilities (CP) over 100 repeated experiments. Prior configurations (ii): $\sigma_\gamma=2, a=b=0.5$.}
    \label{tab:coverage_new_prior}
    \end{table}
Hence, we still choose to use prior setting (i) for the active mediator selection, and only report the metrics (bias, variance, MSE for the fixed-parameters) of point estimation on scalar-valued NIE and NDE (see Tables~\ref{tb:sim_result_NDE}, \ref{tb:sim_result}, and \ref{tb:sim_lowd}).

\section{Sensitivity analysis (SA) in the presence of scalar unobserved confounder}\label{supp_sec:SA_causal}

The proposed BASMU model assumes that the unobserved confounder has a latent structure, and if the unobserved confounder were scalar-valued, it would be inseparable from the random noise $\epsilon_M$ and hence not identifiable. However, in the real data analysis, we never know if there truly exists other one-dimensional unobserved confounder that can impact both the outcome and the mediator. Additionally, although our model allows for the unobserved confounder to be correlated with the exposure $X_i$, this is not directly modeled. In this section, we provide sensitivity analyses assuming that there exist 1-dimensional unobserved confounders, $U_i, V_i ,W_i$, where $U_i$ controls mediator/outcome confounding, $V_i$ controls the treatment/outcome confounding, $W_i$ controls the treatment/mediator confounding. We vary the corresponding effect sizes $\rho_y,\rho_m$ for $U_i$ on the outcome and mediator, respectively. Similarly, we vary the size of $\rho_{vy},\rho_{wm}$ for the effect of $V_i$ on the outcome, effect of $W_i$ on the mediator, respectively. We let $\rho_{vx}=\rho_{wx}=0.1$ to keep the total number of combinations in control.

The equivalent models for BASMU \eqref{eq:full} and \eqref{eq:mediator} in this simplified sensitivity analysis setting are as follows,
\begin{align*}
M_i(s_j) &= \alpha(s_j) X_i + \sum_{k=1}^q \xi_k(s_j) C_{i,k} + \eta_i(s_j)+ \rho_m(s_j) U_i +\rho_{wm}(s_j) W_i + \epsilon_{M,i}(s_j), \numberthis\label{eq:U_m}\\
Y_i  &=\sum_{j=1}^p \beta(s_j)M_i(\Delta s_j)  + \gamma X_i + \sum_{k=1}^q\zeta_k C_{i,k} + \sum_{j=1}^p \nu(s_j)\eta_i(s_j)\leb(\Delta s_j) +\\
&\rho_{vy} V_i+  \rho_y U_i+ \epsilon_{Y,i}, \numberthis\label{eq:U_y}\\
    X_i &= \rho_{vx} V_i + \rho_{wx} W_i +\epsilon_{X,i},\numberthis\label{eq:U_x}\\
    U_{i}, &\sim G(u;\theta_U),  V_i\sim G(v;\theta_V), W_i \sim G(w;\theta_W). \numberthis\label{eq:U}
\end{align*}
where $\epsilon_{X,i}\overset{\iid}{\sim} N(0,\sigma_X^2)$, $\epsilon_{Y,i}\overset{\iid}{\sim} N(0,\sigma_Y^2)$, $\epsilon_{M,i}(s_j)\overset{\iid}{\sim} N(0,\sigma^2_M)$.

We take the mediator/outcome confounder $U_i$ as an example to explain the SA model, and the explanation for $V_i,W_i$ follows similarly. Here, we assume the sampling distribution of the scalar-valued unobserved confounder $U_i$ is $G(u;\theta_U)$. Because there can be infinite combinations of spatially varying effect of $U_i$ on $M_i(s)$, we only consider the spatial homogeneous effect, i.e., assuming the effect of $U_i$ on $M_i(s)$ is $\rho_m(s)\equiv \rho_m$ everywhere. If the sampling distribution $G(u;\theta_U)$ of $U_i$ is allowed to be any general 1-dimensional distribution, the identifiability of the joint model \eqref{eq:U_m} - \eqref{eq:U} is not guaranteed. Hence, we follow a simpler SA model proposed by \cite{Dorie2016-om} where $\bfU_i$ is a single binary variable taking values in $\br{0,1}$, and propose the following SA Algorithm~\ref{algo:sa} for BASMU.

\begin{algorithm}[ht]
\caption{SA algorithm for BASMU with a single binary unobserved confounder}
\label{algo:sa}
\begin{algorithmic}[1]
\State Based on the joint model \eqref{eq:U_m} - \eqref{eq:U} where $U_i\sim \text{Ber}(p_u)$, $V_i\sim \text{Ber}(p_v)$, $W_i\sim \text{Ber}(p_w)$, and assume a spatial constant effect of $\rho_m(s_j),\rho_{wm}(s_j)$ such that $\rho_m(s_j)\equiv\rho_m$, $\rho_{wm}(s_j)\equiv\rho_{wm}$ for all $j$.
\State Specify a combination of fixed values $\rho_y,\rho_m,\rho_{vx},\rho_{wx},\rho_{vy},\rho_{wm}$.
\For {each combination of fixed  $\rho_y,\rho_m,\rho_{vx},\rho_{wx},\rho_{vy},\rho_{wm}$}:
\State Initialize $U_i,V_i,W_i$ to 0. Initialize all other parameters in BASMU.
    \For {each MCMC iteration $t$}:
    \State Draw all BASMU parameters $\bftheta_{\text{BASMU}}$ based on \eqref{eq:U_m} - \eqref{eq:U_x} with fixed $\rho_y,\rho_m,\rho_{vx},\rho_{wx},\rho_{vy},\rho_{wm}$.
    \State Draw $U_i$ independently for each $i$ conditional on the joint model \eqref{eq:U_m} - \eqref{eq:U_x} 
    \begin{align*}
        &U_i \sim \text{Ber}(\pi_{u1}/\sbr{\pi_{u0}+\pi_{u1}})\\
        &\pi_{u1} = p_y(Y_i|U_i=1,\bftheta_{\text{BASMU}}) p_m(\bfM_i|U_i=1,\bftheta_{\text{BASMU}})p_u\\
        &\pi_{u0} = p_y(Y_i|U_i=0,\bftheta_{\text{BASMU}}) p_m(\bfM_i|U_i=0,\bftheta_{\text{BASMU}})(1-p_u)
    \end{align*}
    $V_i,W_i$ are drawn based on the joint model in a similar way. $\pi_{v1} = p_y(Y_i|V_i=1,\bftheta_{\text{BASMU}}) p_x(X_i|V_i=1,\bftheta_{\text{BASMU}})p_v$. $\pi_{w1} = p_m(\bfM_i|W_i=1,\bftheta_{\text{BASMU}}) p_x(X_i|V_i=1,\bftheta_{\text{BASMU}})p_w$.
    \EndFor
    \State Output the posterior distribution of $\bftheta_{\text{BASMU}}$.
\EndFor
\end{algorithmic}
\end{algorithm}

Note that due to the computational constraints, it is challenging to sample $\eta_i$ jointly with all other parameters. In this SA, we keep $\eta_i$ fixed at the same values as in the Two-stage analysis, and use the posterior means of the Two-stage parameters presented as main results in Section~\ref{sec:real_data} as the initial values in the SA. We run a total of $2000$ iterations and take the last 200 MCMC samples as the posterior sample for computing the posterior means of NIE and NDE.

\begin{figure}[ht]
    \centering
    \centering
\begin{subfigure}{.45\textwidth}
  \centering
  \includegraphics[width=\linewidth]{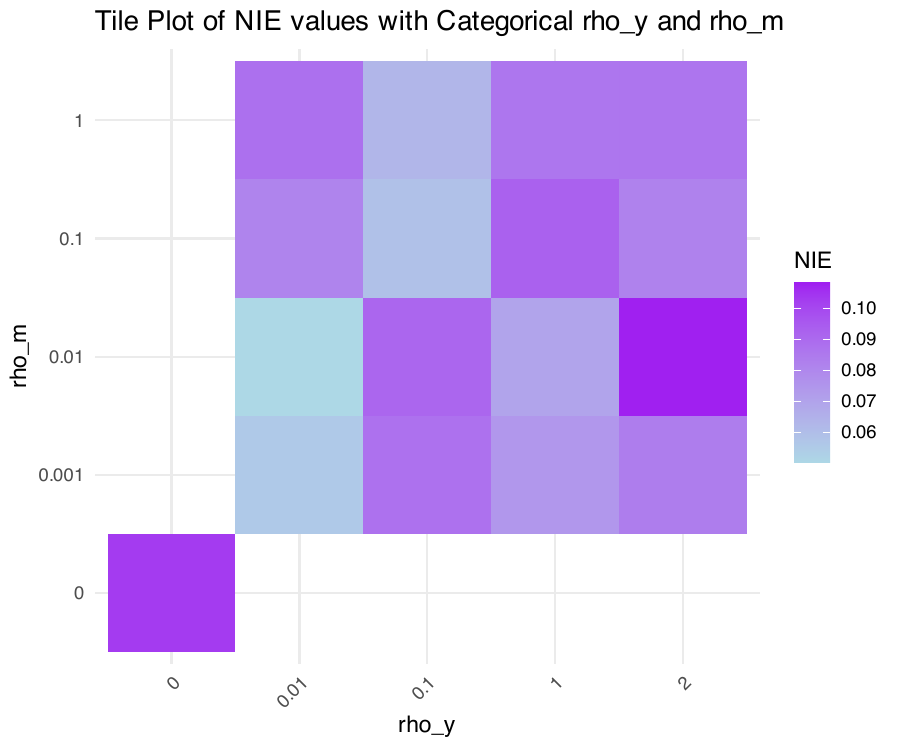}
    \caption{SA 1: Posterior mean of NIE under varying $\rho_y$ and $\rho_m$}
    \label{fig:SA_NIE}
\end{subfigure}%
\begin{subfigure}{.45\textwidth}
  \centering
  \includegraphics[width=\linewidth]{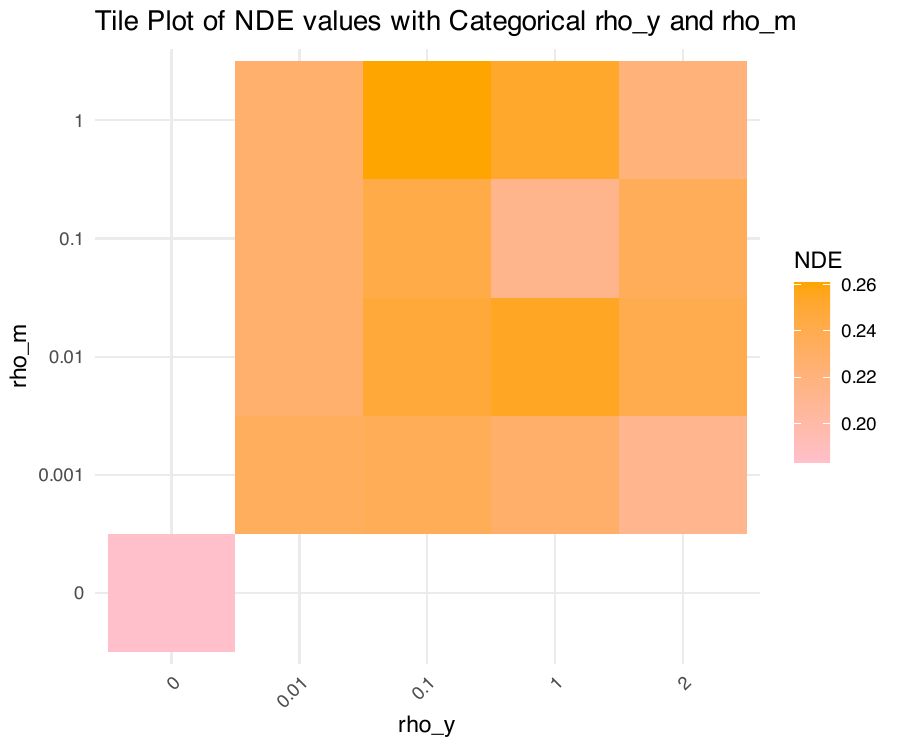}
    \caption{SA 1: Posterior mean of NDE under varying $\rho_y$ and $\rho_m$}
    \label{fig:SA_NDE}
\end{subfigure}

\begin{subfigure}{.45\textwidth}
  \centering
  \includegraphics[width=\linewidth]{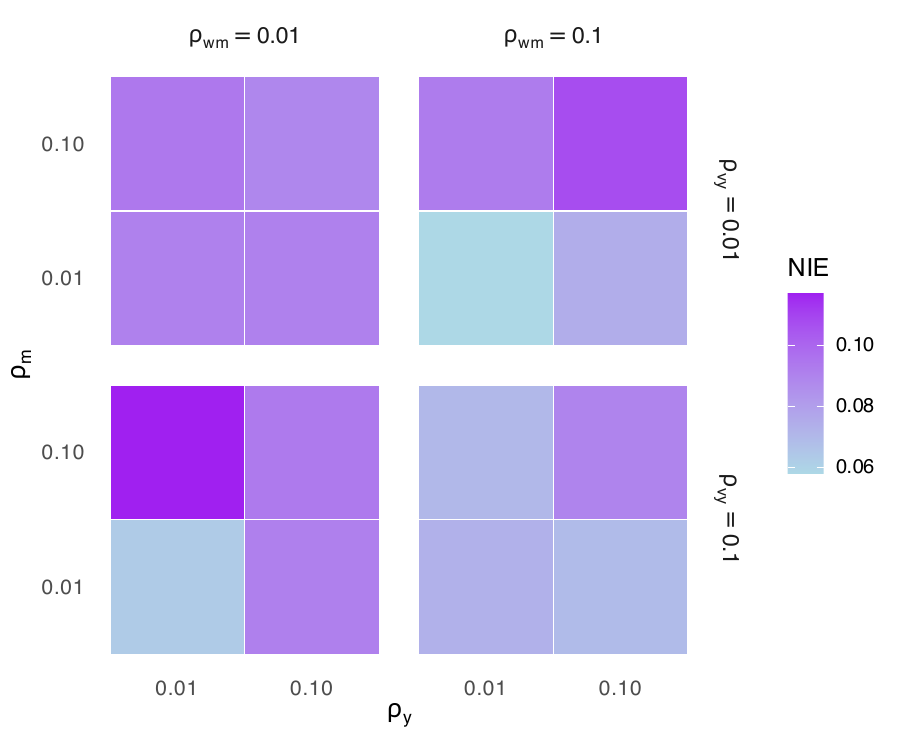}
    \caption{SA 2: Posterior mean of NIE under varying $\rho_y,\rho_m,\rho_{vy},\rho_{wm}$.}
    \label{fig:SA_NIE_full}
\end{subfigure}%
\begin{subfigure}{.45\textwidth}
  \centering
  \includegraphics[width=\linewidth]{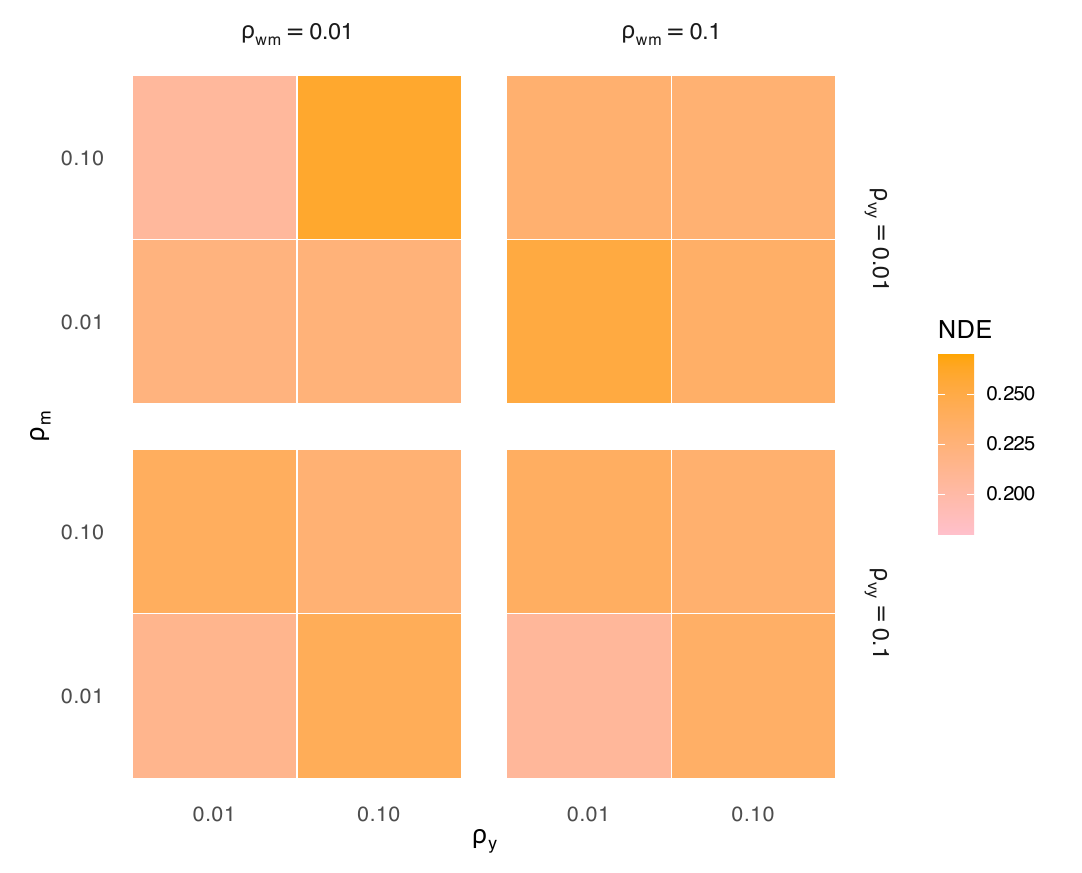}
    \caption{SA 2: Posterior mean of NDE under varying $\rho_y,\rho_m,\rho_{vy},\rho_{wm}$.}
    \label{fig:SA_NDE_full}
\end{subfigure}

\caption{Sensitivity analysis result in the presence of a binary unobserved confounder.}
\label{fig:SA}
\end{figure}

In the first SA (SA 1), we let $\rho_{vx}=\rho_{wx}=\rho_{vy}=\rho_{wm}=0$. Figure~\ref{fig:SA_NIE} and Figure~\ref{fig:SA_NDE} shows the sensitivity analysis result for the combinations of $\rho_m\in \sbr{0.001,0.01,0.1,1}$ and $\rho_y\in \sbr{0.01,0.1,1,2}$. In the context of the ABCD study, the binary unobserved confounder could mean whether the children's nutrition intake or exercise level meets a certain criterion or not. 
In the second SA (SA 2), we let $\rho_{vx}=\rho_{wx}=0.1$, and vary all other effect sizes $\rho_y,\rho_m,\rho_{vy},\rho_{wm}\in\br{0.01,0.1}$. The posterior means of NIE and NDE under all combinations of unobserved effects are shown in Figure~\ref{fig:SA_NIE_full} and Figure~\ref{fig:SA_NDE_full}. Based on the results in Figure~\ref{fig:SA}, we can see that the NIE and NDE effects show small variations around the main result estimates where $\rho_m=\rho_y=\rho_{vx}=\rho_{wx}=\rho_{vy}=\rho_{wm}=0$ (bottom left corners of Figure~\ref{fig:SA_NIE} and Figure~\ref{fig:SA_NDE}).

\section{Sensitivity analysis under different kernel specifications}\label{supp_sec:SA_ker}

In the main analysis, we use a region-specific Mat\'ern kernel where the smoothness parameters differ for each region. As discussed in Section \ref{sec:method}, the latent structure of the subject-specific spatial confounding effect $\eta_i(s)$ matters in the real data analysis. The kernel we used in the main analysis is chosen based on smoothness parameters that roughly match the empirical correlation, and slightly tuned to be more flexible. Since we don't know the ground-truth latent structure of $\eta_i(s)$, in this section, we provide sensitivity analysis results for various kernels. 

According to the Mat\'ern kernel formulation in \eqref{eq:matern}, the two smoothness parameters are $\tau,\rho$. Initially, we choose a base set of $\tau,\rho$ for each region by empirically matching the GP kernel with the sample covariance matrix. This kernel is referred to as the \textit{base} kernel. Then, we try to set the smoothness parameters $\tau,\rho$ the same across different regions, and we experiment with two such pairs, $(\tau=1,\rho=15)$ and $(\tau=0.2,\rho=80)$. Lastly, we also try to vary the \textit{base} kernel a little by keep $\tau$ the same, and changing $\rho$ to be $90\%$ or $110\%$ of the \textit{base} values of $\rho$ for each region. The $110\%\rho$ based on the  \textit{base} kernel is the GP kernel we used in the main analysis in Section~\ref{sec:real_data}.

To evaluate the performance of different kernels, we randomly split the whole data set into 50\% training and 50\% testing, and compute the test MSE for different kernels. The latent structure of $\eta_i(s)$ also plays a key role in the mediation analysis; hence, we include two sets of sensitivity analysis over kernels: (i) using the same $\eta_i(s)$ as in our main analysis and only vary the kernel for $\beta(s)$'s GP prior; (ii) mapping the main analysis estimated $\eta_i(s)$ into different low-dimensional spaces using each kernel listed above, and mapping it back to the original dimensions. Experiment (i) demonstrates the case when $\beta(s)$ and $\eta_i(s)$ do not share the same low-dimensional space. Experiment (ii) demonstrates the case where $\eta_i(s)$ are estimated in different RKHS (reproducing kernel Hilbert space), i.e., having different latent structures. We use the posterior means of the main analysis as the initial values, and run 1000 MCMC iterations with the last 20\% as the MCMC sample after burn-in.

Table~\ref{tb:Ker_SA} provides the test MSE across different kernels. For Experiment (i), where the kernel for $\eta_i(s)$ and $\beta(s)$ differs, with the same $\eta_i(s)$, the last kernel $110\%\rho$ has the lowest test MSE. This is also the kernel used in our main analysis, indicating that for the estimated $\eta_i(s)$, the $110\%\rho$ is the best kernel to use for the GP prior of $\beta(s)$ among all other competitors. For Experiment (ii), where $\eta_i(s)$ is mapped into different RKHS but $\beta(s)$ uses the same kernel as  $\eta_i(s)$, the first \textit{base} kernel has the lowest MSE, followed by the $110\%\rho$ kernel, indicating that if we force $\beta(s)$ and $\eta_i(s)$ to be in the same RKHS, the \textit{base} kernel may fit the outcome model better.

\begin{figure}[ht]
    \centering
    \centering
\begin{subfigure}{.5\textwidth}
  \centering
  \includegraphics[width=\linewidth]{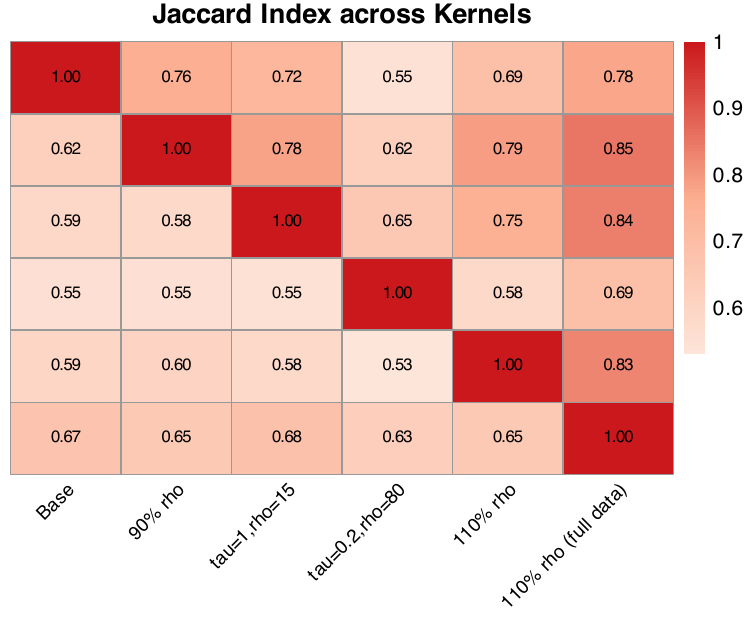}
    \caption{Experiment (i): Same $\eta_i$}
    \label{fig:Ker_SA_sameGP}
\end{subfigure}%
\begin{subfigure}{.5\textwidth}
  \centering
  \includegraphics[width=\linewidth]{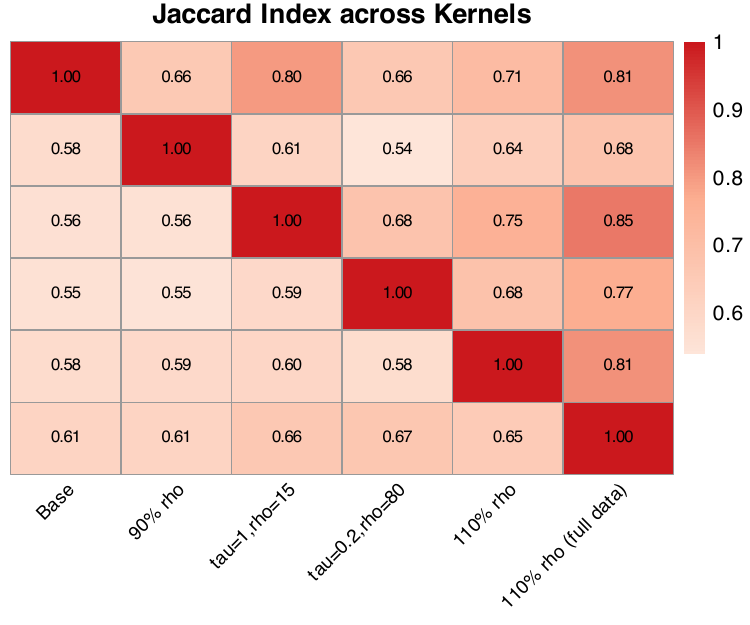}
    \caption{Experiment (ii): $\eta_i$ mapped to different RKHS}
    \label{fig:Ker_SA_diffGP}
\end{subfigure}
\caption{Sensitivity analysis result under different kernel specifications. The numbers in the heatmap represent the values of the Jaccard Index ($\frac{|A\cap B|}{|A\cup B|}$) for the similarity of identified active voxels for each pair of different kernel specifications. Columns and Rows 1 to 5 correspond to the 5 kernrel specifications in Table \ref{tb:Ker_SA}, based on the training data. Column and Row 6 is the main analysis in Section~\ref{sec:real_data} using the $110\%\rho$ kernel based on the entire data set. Numbers in the top right triangular matrix are Jaccard Indices for voxels with $|\cE(s)|>10^{-3}$, based on the main analysis, and the numbers in the bottom right triangular matrix are  Jaccard Indices for voxels with $|\cE(s)|>10^{-4}$.}
\label{fig:Ker_SA}
\end{figure}

In addition to the model fitting result, we are also interested in the active mediation voxel selection. Figure \ref{fig:Ker_SA} provides the Jaccard Index values for similarity comparisons of the active voxels identified by different kernel configurations. For index sets $A$ and $B$ (the index set for active voxels in our case), the Jaccard Index is a measure for similarity, defined as $\frac{|A\cap B|}{|A\cup B|}$. The active voxel selection among different kernels tends to differ on voxels with very low effect sizes, hence we report the similarity results in two levels of voxels with reasonably large effect sizes. The top right triangular matrix in Figure \ref{fig:Ker_SA} is the similarity on voxels with $|\cE(s)|>10^{-3}$ (as estimated in the main analysis). The bottom left triangular matrix is the similarity on voxels with $|\cE(s)|>10^{-4}$. We also included the main analysis selection in Column(Row) 6 in both matrices in Figure \ref{fig:Ker_SA}. Columns (Rows) 1 to 5 are based on training data, but with different kernel settings, and Column(Row) 6 is the main analysis based on the full data. 

Based on Figure \ref{fig:Ker_SA}, when thresholded at $|\cE(s)|>10^{-3}$, the similarities are higher across all pairs. Within the upper triangular matrices, the kernel $\tau=0.2,\rho=80$ has the lowest selection similarity with all other kernels, and this is indicated in Table \ref{tb:Ker_SA} as well because this is the kernel with large test MSEs. The last column, the main analysis result, has the highest similarity with all other kernels since this result is based on the full data analysis. If we only compare Columns (Rows) 1 to 5, all based on training data alone, on the left panel with same $\eta_i(s)$, all kernels except for $\tau=0.2,\rho=80$ have good selection similarity with Jaccard Index around or above 0.7. When $\eta_i(s)$ belongs to different RKHS, on the right panel, the similarity is generally lower, although the trio: \textit{base} kernel, the $\tau=1,\rho=15$ kernel, and the $110\%\rho$ kernel still have pairwise Jaccard Index greater than 0.7. 

This sensitivity result shows that, although changing the kernel structure for $\eta_i(s)$ or $\beta(s)$ will have impact on both the model fitting and the active mediation voxel selection, there still exist certain common low-dimensional structures that maintain over 70\% selection similarity to be consistent across different kernel specifications and random subsetting of the full data. Although the current kernel we use may not be the best kernel that truly reflects the underlying low-dimensional structure of the unobserved confounder, it is still a practically stable choice.

\begin{table}[ht]
\centering
\caption{Summary of test MSE for different kernels.}
\label{tb:Ker_SA}
\begin{tabular}{lccccc}
\toprule
 & Base & $90\%\rho$ & $\tau=1,\rho=15$ & $\tau=0.2,\rho=80$ & $110\% \rho$ \\
\hline
test MSE (i) & 0.652 & 0.728 & 0.689 & 0.724 & 0.616 \\
test MSE (ii)  & 0.601 & 0.643 & 0.747 & 0.733 & 0.621 \\
\bottomrule
\end{tabular}
\end{table}

\section{Additional discussion on the limitation of BASMU}\label{supp_sec:additional_limitation}

The limitation of this work is that the mediator has to be spatially smooth or satisfy certain pre-fixed correlation structures for the subject-specific spatial confounding effects to be estimated. In practice, unobserved confounders with more complex correlation structures may not be fully accounted for under the current BASMU framework.

As discussed in \cite{vanderweele2014effect}, the \textit{cross-world independence assumption} (assumption (iv)  in \eqref{eq:unobserved_assumption}) requires that, in the true data generating mechanism, there is no post-exposure latent variables, affected by $X_i$, that also affect both the mediator $M_i(s)$ and outcome $Y_i$, while not being part of the mediator of interest (i.e., no causal direction $X_i\to\eta_i$ while $\eta_i\to M_i, \eta_i\to Y_i$ at the same time). This assumption is an untestable causal identification assumption, especially in the BASMU model, where $\eta_i$ is an individual-level latent effect. In the real data analysis, $X_i$ is the parental education level, and $\eta_i$ is the spatial-varying individual effect coming from the mediator model, after regressing out $X_i$ and $\bfC_i$. In BASMU, $\eta_i(s)$ is intended to represent subject-specific latent heterogeneity outside the targeted mediation pathway. A few examples of such subject-specific spatial effects include: i) genetic polymorphisms influencing cortical maturation, since genetics are fixed at birth, clearly not caused by parental education; neurotoxic pollutant exposure (lead, air pollution), which is largely determined by residential environment, plausibly independent of parental education level;
baseline sleep or physical activity patterns, which reflect stable physiological and behavioral traits rather than parental behavior.  \label{sent:limitation}

The focus of BASMU is to provide a way to reduce bias in the point estimation of scalar-valued NIE and NDE, and to select active mediation voxels using the criterion whether the Bayesian 95\% credible intervals cover 0. The frequentist-style inference capability, such as the frequentist coverage probabilities of the point estimates for NIE and NDE in this Bayesian framework, would depend on the choice of the priors, especially priors for the variance parameters $\sigma_\gamma^2,\sigma_\beta^2,\sigma_\nu^2$. To achieve such frequentist coverage probabilities in finite samples, one needs to carefully set the priors to be adaptive to the sample size, which is beyond the scope of the current BASMU framework.

\end{document}